\documentclass[11pt,a4paper]{article}
\pdfoutput=1 
\usepackage{jheppub}	

  \usepackage{tikz}
\usetikzlibrary{positioning,arrows}
\usetikzlibrary{decorations.pathmorphing}
\usetikzlibrary{decorations.markings}
\usetikzlibrary{snakes}
\usetikzlibrary{matrix}

\usepackage[utf8]{inputenc}
\usepackage[english]{babel}
\usepackage{makeidx}
\usepackage{tikz}
\tikzset{every picture/.style={line width=0.75pt}} 
\usepackage{amsfonts}
\usepackage{enumerate}
\usepackage{mathrsfs}
\usepackage{tensor}
\usepackage[autostyle]{csquotes}
\usepackage{subfig}

\def\be{\begin{equation}}
\def\ee{\end{equation}}
\def\bea{\begin{eqnarray}}
\def\eea{\end{eqnarray}}
\newcommand{\nn}{\nonumber}

\newcommand{\ft}[2]{{\textstyle\frac{#1}{#2}}}

\def\apjl{\ref@jnl{ApJ}}

\usepackage{amsmath,amssymb}
\usepackage{latexsym}
\usepackage{graphicx}
\usepackage{slashed}

 \usepackage[vcentermath,enableskew]{youngtab}
\usepackage{epsfig}
\newcommand{\tr}{\mathrm{tr}}

\def\be{\begin{equation}}
\def\ee{\end{equation}}
\def\bea{\begin{eqnarray}}
\def\eea{\end{eqnarray}}

\def\cF{\mathcal{F}}

\title{CFT description of BH's and ECO's:  QNMs, superradiance, echoes and tidal responses}

\author[a]{Dario Consoli,}
\author[a]{Francesco Fucito,}
\author[a]{Jose Francisco Morales,}
\author[b]{Rubik Poghossian.}
\affiliation[a]{Dipartimento di Fisica, Università di Roma ``Tor Vergata"  \& Sezione INFN Roma2, Via della ricerca
scientifica 1, 00133, Roma, Italy}
\affiliation[b]{Yerevan Physics Institute,
Alikhanian Br. 2, 0036 Yerevan, Armenia}
\abstract{ Using conformal field theory and localization tecniques we study the propagation of scalar waves in gravity backgrounds described by Schr\"odinger like equations with Fuchsian singularities. Exact formulae for the connection matrices relating the asymptotic behaviour of the wave functions near the singularities are obtained in terms of braiding and fusion rules of the CFT.  The results are applied to the study of quasi normal modes, absorption cross sections, amplification factors, echoes and tidal responses of  black holes (BH) and exotic compact objects (ECO) in four and five dimensions. In particular, we propose a definition of dynamical Love numbers in gravity.}

\newcommand{\commentfr}[1]{{\color{red} #1}}

 \clearpage
\begin{document}
\tikzset{
line/.style={thick, decorate, draw=black,}
 }

\maketitle

\flushbottom

\section{Introduction }
Very recently we have entered the era of gravitational wave (GW) astronomy with the first measurements  coming from
coalescing binaries \cite{LIGOScientific:2016aoc}.  Can this experiments test fundamental physics?  The coalescence of two astrophysical objects can be divided into three phases: the inspiral stage where the two objects are well separated  and orbit around each other,  the merger phase when they coalesce and the ringdown phase when the  compact remnant relaxes emitting gravitational waves of very characteristic frequencies, the so called quasi-normal modes (QNM's) \cite{Berti:2009kk}. The   inspiral and  ring-down phases can be studied using space-time perturbation theory. Indeed, the tidal response of a neutron star can be traced from the GW signal   emitted in the late inspiral stage  and expressed in terms of Love numbers \cite{Flanagan:2007ix}. On the other hand, QNM's dominate the ring-down phase.

 Can the GW signal be used to distinguish BH's from exotic objects more massive than neutron stars but without a horizon ?
The list of alternatives is long: gravastars, wormholes, firewalls and fuzzballs,  collectively known as Exotic Compact Objects (ECO) \cite{Cardoso:2019rvt}.  ECO's tend to be as black as BH's. If compact enough, they are typically surrounded by a photon sphere where  both massive and massless probes orbit in unstable circular motion \cite{Claudel:2000yi}.  External perturbations excite the QNM's which appear as damped vibrations of the ECO response.
A fraction of this signal, moving towards the ECO interior, is reflected back producing a sequences of echoes that can be viewed as the smoking gun for the existence of a horizonless object \cite{Cardoso:2017cqb}.  Several attempts to distinguish BH's from fuzzballs studying their photon sphere sizes and shapes, QNM's, multipolar structure and echoes  has recently appeared in \cite{Bianchi:2018kzy}\nocite{Bianchi:2020des,Bena:2020see,Bianchi:2020bxa,Bena:2020uup,Bianchi:2020miz,Mayerson:2020tpn,Bianchi:2020yzr,Ikeda:2021uvc}-\cite{Bah:2021jno}.
Hidden symmetries of black hole and fuzzball photon spheres have been explored in \cite{Bianchi:2021yqs,Bianchi:2022wku}. Echoes produced by black holes in modified theories of gravity have been studied in \cite{Dong:2020odp}.

    In this paper, we study the wave response of BH's and ECO's to small deformations of the geometry.  These studies were pioneered by Regge and Wheeler in the late 50's \cite{Regge:1957td}.  The wave propagation in  the Schwarzschild geometry  turns out to be a Schr\"odinger like equation describing the scattering of a particle in the presence of an effective potential. The results were generalized  by Teukolski to the case of Kerr-Newman geometry
    \cite{1972ApJ...178..347B,Chandrasekhar:1985kt,Novikov:1989sz}.
 In general, these equations cannot be solved analytically even for very simple BH's, so most of the available results rely on numerical analysis.

     Quite recently, a new astonishing connection between the QNM spectral problem and the quantum periods of the Seiberg-Witten (SW) geometries was proposed in \cite{Aminov:2020yma} and tested in the Kerr case. The correspondence was extended in
    \cite{Bianchi:2021xpr,Bianchi:2017bxl} to several gravity systems including BH's in higher dimensions, D-brane bound states and fuzzballs. Moreover, using the AGT correspondence \cite{Alday:2009aq}, the solution of the wave equation in the Kerr case  was identified with certain five-point conformal blocks of a two-dimensional conformal field theory (CFT). This provides a new tool to study other interesting observables such as Love numbers, absorption coefficients and grey body factors \cite{Bonelli:2021uvf}.  QNM's and connections matrices have been also determined in \cite{Fioravanti:2021dce} using techniques based on integrability.

     In this work, following \cite{Bonelli:2021uvf}, we apply CFT and localization techniques to the study of  wave equations describing radial and angular motion in separable gravity backgrounds described by Schr\"odinger like equations with Fuchsian singularities.
      We relate the solutions of a differential equation with $N$ Fuchsian singularities to  $(N+1)$-point conformal blocks of Liouville theory  involving the insertion of $N$ primaries and a degenerate field, i.e. a primary field containing a null state in its Verma module. These conformal blocks obey a differential equation known as (BPZ)\cite{Belavin:1984vu} and are related via AGT correspondence \cite{Alday:2009aq} to ${\cal N}=2$ supersymmetric quiver gauge theories living in curved space times. In a particular $\Omega$-background (NS) \cite{Nekrasov:2009rc}, the  gauge dynamics is given by a quantization of the SW curve \cite{Poghossian:2010pn,Fucito:2011pn} and the BPZ equation reduces to  an ordinary differential equation that  can be identified with the wave equation in gravity.

    Combining the two pictures, we will derive a combinatorial formula for the wave function and exact formulae for the connection matrices relating the asymptotic expansions of the solutions near any two singularities.  These matrices are given by  the "braiding" and ``fusion" rules of the underlying CFT.  The
 results provide a generalization of the Heun connection formulae, recently obtained in \cite{Bonelli:2022ten}, to the general case of ordinary differential equations with arbitrary number of Fuchsian singularities. Solutions of Fuchsian equations have been also studied in  \cite{Novaes:2014lha,CarneirodaCunha:2015hzd,CarneirodaCunha:2015qln,Amado:2017kao,BarraganAmado:2018zpa,Novaes:2018fry,CarneirodaCunha:2019tia,Amado:2020zsr} using isomonodromic methods and  Painlev\`e trascendents.

    We  discuss various applications of the gauge/CFT techniques in the context of BH's and ECO's in four and five dimensions. ECO's are represented by solutions of Einstein equations on space times ending on a membrane with reflectivity properties. The effect of the inner boundary reflectivity coefficient on
   QNM's and echo responses is discussed.  Finally we compute the amplification factors and dynamical Love numbers for slowly oscillating (yet non-static) waves.  For the sake of simplicity, in this paper we restrict ourselves to scalar modes, but all CFT formulae we found apply to higher spin modes as  well, the only difference being in the gauge/gravity dictionary. 

     In the astrophysical realm static Love numbers were first defined and studied in \cite{Damour:2009vw,Binnington:2009bb}.  For four dimensional  Schwarzschild \cite{Fang:2005qq,Damour:2009vw,Binnington:2009bb} and Kerr \cite{Chia:2020yla} BH's  they were shown to vanish.  They are generically  non-trivial in higher dimensions \commentfr{ \cite{Kol:2011vg,Hui:2020xxx,Pereniguez:2021xcj}}, non-asymptotically flat spaces \cite{Emparan:2017qxd} and alternative theories of gravity \cite{Cardoso:2017cfl,Cardoso:2018ptl}.

      For a static wave $\omega =0$, the Love number is defined as the ratio between the coefficients of the terms in the scalar wave function going at infinity as $r^{-\hat{\ell}-1}$ and $r^{\hat{\ell}}$ , with $\hat{\ell}=\ell/(D-3)$. In the rotating case, the exponents of the two solutions get shifted $\hat{\ell}\to \hat\ell+\Delta{\nu}(\omega)$, by the same function of $\omega$.  We define the dynamical Love number as the ratio of the coefficients of the leading terms (with shifted exponents) in the double scaling limit where the two gauge couplings are small and the gauge theory partition function can be computed with localization. In the gravity language this corresponds to the region where  $r_+-r_- << r << \omega^{-1}$ \footnote{Different definitions of the  dynamical Love numbers have been proposed in \cite{Charalambous:2021mea,Bonelli:2021uvf}.}.
   We derive a unique formula describing the Love and dissipation numbers of all BH's of Einsten-Maxwell theory in four and five dimensions in terms of few parameters specifying the gravity background. The entire non-triviality of the differential equation (the Heun equation) is codified into a single function:  the quantum SW period $a(u)$ specifying the gauge theory prepotential ${\cal F}(a)$  ! This function can be derived from a continuous fraction equation representing a quantum deformed version of the SW curve  \cite{Poghossian:2010pn,Fucito:2011pn}.

   This is the plan of the paper: in Section 2 we describe the gauge-CFT-gravity dictionary.  In Section 3 we study the QNM's and echo wave responses of gravity backgrounds on spacetimes ending on a membrane with non-trivial reflectivity properties.  In Sections 4  and 5 we compute amplification factors and dynamical Love numbers for BH's in four and five dimensions using the CFT description of the wave functions and connection matrices.  In Section 6 we summarise our results.

\section{SW Gravity correspondence}

In this section, we review the SW gravity correspondence in its most general set up.  We consider a gravity system whose wave equation can be separated into
 ordinary differential equations with Fuchsian singularities describing radial and angular motion.
 An  ordinary differential equation of second order with $N$ Fuchsian singularities can be always put into the normal form
\be
\left[ {d^2\over dx^2} +Q(x) \right] \Psi(x) = 0  \label{can}
\ee
with
\be
Q(x)=\sum_{i=1}^N \left[ {\delta_i \over (x-x_i)^2} + {c_i \over (x-x_i) }\right] \label{q0}
\ee
where $\delta_i$, $c_i$, $x_i$  are some constants parametrizing the positions $x_i$ of the singularities and the behaviour of the equation in their neighborhood.
Here we assume that infinity is a regular point, i.e. $Q(x)$ vanishes at $x^{-4}$, at large $x$,  i.e.
\be
\sum_{i=1}^N c_i =\sum_{i=1}^N (c_i\, x_i+\delta_i)=\sum_{i=1}^N (c_i \,x^2_i+2\,\delta_i \,x_i )=0 \label{inf3}
\ee
 There is a coordinate freedom in choosing the location of the singularities. We will always use this freedom to  map the locations of three of the singularities $x_0,x_1,x_\infty$ to $0,1,\infty$.
We take
\be
x \to  z(x) ={ x-x_0\over x-x_\infty} {x_1-x_\infty\over x_1-x_0}
\ee

 Under a general coordinate transformation $x\to  z(x)$,
  a differential equation of the form (\ref{can}) can be brought to its normal form by taking
  \be
 Q( z)=Q(x( z) ) \, x'( z)^2 +{ x'''(  z) \over 2  x'( z) } -\frac{3}{4} \left[ {  x''(  z) \over  x'(  z) }
\right]^2 \quad ,\quad    \Psi( z) = { \Psi( x( z) ) \over \sqrt{ x'( z)} }
\label{qqschwarzian}
\ee
where the prime stands for derivation. (\ref{inf3}) can be solved for $c_0$, $c_1$ and $c_\infty$  as
 \be
 c_{\infty }  =0 \quad , \quad   c_{1} =\delta {-} \sum^N_{i \neq 0,1,\infty}  z_i c_i    \quad , \quad
c_0   =-\delta +  \sum^N_{i \neq 0,1,\infty}(  z_i-1) c_i
 \ee
 with
 \be
 \delta=2\delta_\infty -\sum_{i=1}^N  \delta_i
 \ee
 The aim of this section is to derive a combinatorial formula for the solutions $\Psi_\alpha (z)$  and the connection matrices relating its asymptotic behaviour
near the singularities.

 \subsection{The CFT description}

  We consider a Liouville theory
with background charge $Q$ and central charge $c$ given by
\be
c=1+6\,Q^2 \quad, \quad  Q=  b+{1\over b}
\ee

 \subsubsection{The conformal blocks}

 We consider the $(n+3)$-point correlator involving the insertion of chiral primary fields with momenta
 \be
 P_s=\ft{1}{b}(p_0,k_0,k_1\ldots k_n,p_{n+1})
 \ee
 and vertices  $V_{P_s}=e^{(Q-2P_s) \phi} $. We denote by
   \footnote{
As usual, we define $\langle   p_0 |   =\underset{z_{-1} \to \infty}{\lim}   \langle  0 | z_{-1}^{2\Delta_{p_0} } V_{p_0} (z_{-1} ) $ and $| p_{n+1}\rangle= V_{p_{n+1}}(0 ) |0\rangle $  }
\be
 \cF_{p_0}{}^{k_0}{}_{p_1} .. {}^{k_{n}}{}_{p_{n+1} }   (x_i)
   =     \left \langle   p_0 |  V_{k_0}(x_0)  \ldots V_{k_{n} }(x_{n} )  | p_{n+1}  \right\rangle _{p_1\ldots p_n} =
 \begin{tikzpicture}[scale=0.8,baseline={([yshift=-.5ex]current bounding box.center)}]
	\draw [] (11,0)--(13.2,0);
	\draw [] (13.4,0)--(13.5,0);
	\draw [] (13.6,0)--(14.1,0);
	\draw [] (14.1,0)--(15.1,0);
	\node [left] at (11,0) {\scriptsize{$\infty$}};
	\node [below] at (11.6,0) {\scriptsize{$ p_0$}};
	\node [below] at (12.6,0) {\scriptsize{$p_{1}$}};
	\node [below] at (14,0) {\scriptsize{$p_{n}$}};
	\node [below] at (14.95,0) {\scriptsize{$ p_{n+1}$}};
	\node [right] at (15.1,0) {\scriptsize{$0$}};
	\node [left] at (12,0.7) {\scriptsize{$k_0$}};
	\node [above] at (12,1) {\scriptsize{$x_0$}};
	\draw [] (12,0)--(12,1);
	\draw [] (13,0)--(13,1);
	\node [left] at (13,0.7) {\scriptsize{$k_1$}};
	\node [above] at (13,1) {\scriptsize{$x_1$}};
	\draw [] (14.3,0)--(14.3,1);
	\node [left] at (14.3,0.7) {\scriptsize{$k_n$}};
	\node [above] at (14.3,1) {\scriptsize{$x_n$}};
	 	\end{tikzpicture}
\ee
the corresponding conformal block with $p_1, p_2, \ldots p_n$ specifying the intermediate momenta.
The dimensions of the operators are related to momenta by
\be
 \Delta_{p_i}={Q^2\over 4} -{p_i^2 \over b^2} \qquad , \qquad   \Delta_{k_i}={Q^2\over 4} -{k_i^2 \over b^2}
\ee

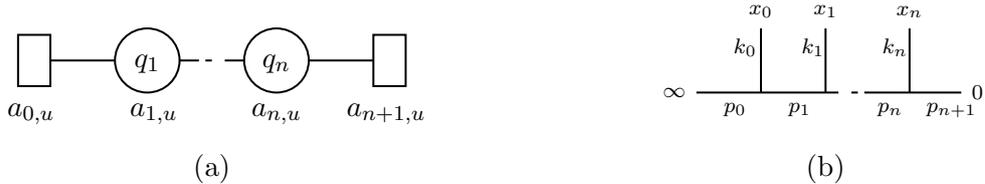
\begin{figure}[t]
	\begin{tikzpicture}[scale=0.85]
	\draw [] (0.5,0.1) rectangle (1,0.9);
	\node [below] at (0.7,0) {$a_{0,u}$};
	\draw [] (1,0.5)--(2,0.5);
	\draw [] (2.5,0.5) circle [radius=0.5];
	\node [] at (2.5,0.45) {$q_1$};
	\node [below] at (2.6,0) {$a_{1,u}$};
	\draw [] (3,0.5)--(3.3,0.5);
	\draw [] (3.4,0.5)--(3.5,0.5);
	\draw [] (3.7,0.5)--(4,0.5);
	\draw [] (4.5,0.5) circle [radius=0.5];
	\node [] at (4.5,0.45) {$q_n$};
	\node [below] at (4.5,0) {$a_{n,u}$};
	\draw [] (5,0.5)--(6,0.5);
	\draw []  (6,0.1) rectangle (6.5,0.9);
	\node [below] at (6.2,0) {$a_{n+1,u}$};
	\draw [] (11,0)--(13.2,0);
	\draw [] (13.4,0)--(13.5,0);
	\draw [] (13.6,0)--(14.1,0);
	\draw [] (14.1,0)--(15.1,0);
	\node [left] at (11,0) {\scriptsize{$\infty$}};
	\node [below] at (11.6,0) {\scriptsize{$ p_0$}};
	\node [below] at (12.6,0) {\scriptsize{$p_{1}$}};
	\node [below] at (14,0) {\scriptsize{$p_{n}$}};
	\node [below] at (14.95,0) {\scriptsize{$ p_{n+1}$}};
	\node [right] at (15.1,0) {\scriptsize{$0$}};
	\node [left] at (12.1,0.7) {\scriptsize{$k_0$}};
	\node [above] at (12,1) {\scriptsize{$x_0$}};
	\draw [] (12,0)--(12,1);
	\draw [] (13,0)--(13,1);
	\node [left] at (13.15,0.7) {\scriptsize{$k_1$}};
	\node [above] at (13,1) {\scriptsize{$x_1$}};
	\draw [] (14.3,0)--(14.3,1);
	\node [left] at (14.44,0.7) {\scriptsize{$k_n$}};
	\node [above] at (14.3,1) {\scriptsize{$x_n$}};
	\node [below] at (13,-0.8) {(\cal{b})};
	\node [below] at (3.5,-0.8) {(\cal{a})};
	\end{tikzpicture}
	\caption{(\cal{a}) The diagram for the linear quiver $U(2)^n$ gauge theory.
		(\cal{b}) The AGT dual conformal $n+3$-point block of the Toda field theory.}
	\label{quiv_block}
\end{figure}

The conformal block  $\mathcal{F}_{p_0}{}^{k_0}.. {}^{k_{n}}{}_{p_{n+1} }$ is related via the AGT duality to the  instanton partition function
of a $U(2)^n$ quiver gauge theory.  The gauge theory is specified by the scalar vevs $a_{iu}$, with $u=1,2$ a gauge index, $i=0,\ldots n+1$ labelling the quiver nodes, the gauge couplings $q_i$  and the $\Omega$-background parameters $\epsilon_1,\epsilon_2$.
 They are related to the CFT variables  via the AGT dictionary
\be
   p_i =  \frac{a_{i1}-a_{i2} }{2     }   \quad  , \quad
b {Q\over 2}-k_i =  \sum_{u=1}^2  \frac{ a_{i+1,u}-a_{i,u} }{2    }   \quad , \quad
{x_i\over x_{i-1} }  =   q_i
~~,~~ \epsilon_1=b^2 ~,~ \epsilon_2=1
  \label{agtdic}
\ee
 Finally the conformal block is related to the instanton partition function. The latter can be written as a sum over a set of $2n$ Young tableaux $Y=\{ Y_{iu} \}$, $i=1,\ldots n$, while  $Y_{0u}=Y_{n+1,u}=\emptyset$
(see Appendix \ref{sec-inst} for details).
  Translated into the CFT variables, the conformal block can be written as
   \be
 \mathcal{F}_{p_0}{}^{k_0}.. {}^{k_{n}}{}_{p_{n+1} }   (x_i) =  \prod_{i=0}^n   x_{i}^{\Delta_{p_i} {-}\Delta_{p_{i+1} }{-}\Delta_{k_i}}    \prod_{ \underset{i< j}{i,j=0} }^{2}  \left(1{-}{x_{j} \over x_i} \right)^{ (2k_{i}{-}b Q)(2k_{j}{+}b Q) \over 2 b^2  }  \,
 Z_{\rm inst}{}_{p_0}{}^{k_0} .. {}^{k_{n}}{}_{p_{n+1} }   (x_i)   \label{ffinst}
 \ee
 with
 \be
 Z_{\rm inst}{}_{p_0}{}^{k_0} .. {}^{k_{n}}{}_{p_{n+1} }   (x_i)= \sum_{ Y  } \prod_{i=0}^n  x_i^{|\vec Y_{i}| - |Y_{i+1} |}
{    \prod_{u,v}^2    {\cal Z}_{Y_{i u},Y_{i+1,v} }(k_i{+} (-)^u p_i {-} (-)^v p_{i+1} -b \ft{Q}{2}  )
 \over        {\cal Z}_{Y_{i1},Y_{i2} }( 2 p_i ) {\cal Z}_{Y_{i2},Y_{i1} }( {-}2 p_i )} \label{zinstG}
 \ee
 and
 \be
{\cal Z}_{Y_u,W_v}(y)
= \prod_{i,j\in Y_u}  \left[ y{+}\epsilon_1 (i{-}{k}_{jv}'){-} \epsilon_2 ( j{-}{k}_{iu}{-}1) \right]  \prod_{i,j\in W_v} \left[ y  {-}\epsilon_1 (i{-}{k}_{ju}'{-}1){+} \epsilon_2 ( j{-}{k}_{iv})   \right] \label{zyy}
\ee
  Here $i,j$ run over the columns and rows of the tableau and $k_{iu}$,$k'_{ju}$ denote the height and lengths of the columns and rows of the tableau $Y_u$.

 The first product in (\ref{ffinst}) in front of $Z_{\rm inst}$ can be viewed as the tree level contribution to the quiver partition function. The second one subtract the U(1) part.  In principle one could also include a one-loop contribution  but since it is $z_i$-independent, it is irrelevant for our analysis here.
 We notice that neither $z_0$, that accounts for the freedom of an overall rescaling of the vertex field locations, nor the gauge theory center of mass $\sum_{i,u} a_{iu}$ are determined by the AGT dictionary. They can be fixed at will.

\subsubsection{The BPZ equation}

 We are interested in the conformal blocks involving the insertion of a degenerate field. Degenerate fields correspond to primaries with momenta
     \be
     {k_{m m'} \over b} \equiv  {1\over 2b}  m +{b\over 2} m'
\label{degalpha1}
   \ee
     where $m, m'$ are positive integers. To each degenerate field one can associate a null state  with vertex
     \be
V_{\rm null,m m'}={\bf L}_{m m'}  V_{\alpha_{m m'}}
\ee
 where ${\bf L}_{m m'}$ is some polynomial of the Virasoro generators
 at level $N=m\, m'$.  A correlator involving the insertion of a null state vanishes and so the corresponding conformal block will satisfy a differential equation of order $ m m'$.

  In order to get a second order differential equation, we
  consider the descendant at level 2 of $V_{k_{12}}$
  \bea
V_{\rm null,12}&=&\left( L_{-1}^2{+}b^2  L_{-2} \right) V_{k_{12}}   \qquad  k_{12}=\ft{1}{2} + b^2 \qquad \Delta_{12 }={-}\ft12{-}\ft{3 b^2}{4}
\label{vnull}
\eea
 An important simplification comes from the fact that the OPE of a degenerated field $V_{k_{12}}$ with a primary of momentum $p/b$, produces only two operators with momenta \cite{Belavin:1984vu}
 \be
 { p^{\alpha} \over b }  ={p \over b} + \alpha {b \over 2}
 \ee
and $\alpha =\pm$.
 Consequently we denote by
 \be
 \Psi_{ i \alpha} (z_s)=  \mathcal{F}_{p_0}{}^{k_0} ..{}_{p_i^{-\alpha} }{}^{k_{12}}  {}_{p_i} ..{}^{k_n}  {}_{p_{n+1} }(z_s)
 \ee
 the degenerate conformal block obtained by the insertion of a degenerate field at the $i^{\rm th}$ position, i.e. $k_i=k_{12}$.
   The function $\Psi_{ i \alpha} (z_s)$ satisfies the BPZ equation
     \bea
&&  \left[ \partial_{z_i}^2 + b^{2}  \sum_{s \neq i }^{n+3}  \left({\Delta_{P_s}\over (z_i-z_s)^2}{+}
{1\over z_i-z_s}\partial_{z_s}  \right) \right]  \Psi_{i\alpha}   (z_s)  =0
 \label{BPZ}
\eea
     Different choices of $i$ correspond to different orderings of the vertex insertions and provide a different  solution of the same equation.
    On the other hand, given ``i ",  $\alpha=\pm$ labels the two independent solutions of the differential equation.

    To evaluate (\ref{BPZ}) one should first
 use the Ward identities
\be
\label{conformal_Ward_identities}
\sum_{s=1}^{n+3} \partial_{z_s}  =\sum_{s=1}^{n+3}  \left( z_s  \partial_{z_s}  + \Delta_{P_s} \right)   =
\sum_{s=1}^{n+3}  \left( z_s^2  \partial_{z_s}   +2z_s  \Delta_{P_s} \right)  =0
\ee
  to express  $\partial_{z_s}$ for three of the insertions  in terms of the remaining ones, and then set the three positions to $\infty,1,0$.
  The resulting equation involves derivatives with respect to $z$ and the remaining $x_s$.
    To reduce it to an ordinary differential equation in the variable $z$, one takes the $\Omega$-background with $\epsilon_1\neq 0, \epsilon_2=0$  \cite{Nekrasov:2009rc}. This corresponds to the limit
  \be
   b\to 0  \qquad  , \qquad \Delta_s \to \infty \qquad, \qquad    \delta_s=b^2 \Delta_s :{\rm finite}
 \label{NSlimit}
 \ee
   for $s\neq  i$ while  $  b^2 \Delta_{i} \to 0$. Notice that in this limit  $p_s$ and $k_s$ are finite.
    An explicit evaluation of the conformal blocks, denoted generically as  $\Psi(z,x_s)$, shows that in this limit they behave as
  (see for example formula (\ref{zinstexplicit}) for the 5-pt conformal block)
 \be
 \Psi(z,x_s) =e^{ {f_0(x_s) \over b^2} +f_1(x_s,z) +\ldots } \label{f0f1}
 \ee
 with the leading contribution independent of the $z$-variable\footnote{This is not surprising since the dimension $\Delta_{k_{12}}$ of the degenerate field is much smaller that the dimensions of the remaining operators, so its insertion in the four-point correlator cannot modify significantly its leading behaviour. }. Consequently $b^2 \partial_z$-terms  in (\ref{BPZ}) can be discarded while $x_i$-derivatives can be replaced by $z$-independent functions related to the Coulomb branch parameters of the quiver gauge theory with one node less.
 As a result the BPZ equation reduces to an ordinary differential equation of type (\ref{can}) with $N=n+2$ Fuchsian singularities.

\subsection{Braiding and fusion relations}

In this section we determine the relations between the different solutions $\Psi_{i \alpha}$ of the BPZ equations. Throughout this section we keep $b$ finite.

 To establish the braiding and fusion relations it is enough to consider the simplest $n=1$ case, i.e. the case of a degenerate four-point conformal blocks. The corresponding BPZ equation in this case contains three Fuchsian singularities and therefore can be explicitly solved in terms of hypergeometric functions. The BPZ equation reads
      \be
\label{BPZeq4}
\left(  b^{-2} \partial_z^2   {+} \frac{1{-}2 z}{z (z{-}1)}\,\partial_z {+}\frac{ \Delta_{p_0}- \Delta_{k}-\Delta_{12} - \Delta_{p_2}   }{z(z{-}1)}{+}  \frac{\Delta_{p_2} }{z^2}{+}  \frac{\Delta_{k} }{(z{-}1)^2}
\right) \Psi  (z)=0\,
\ee
This equation can be brought to the standard hypergeometric form by writing
\be
\Psi(z)=(1-z)^{\left( {1\over 2}+k_0+{b^2\over 2}\right)} z^{\left( {1\over 2}+p_2+{b^2\over 2}\right)} F(z) \label{eq4}
\ee
with $F(z)$ a hypergeometric function. There are six natural choices for $F(z)$ denoted by $w_i$ in \cite{822801}. The six functions $w_i$ can be written as (up to constants )
\bea\label{allsolutions}
\Psi^R_\alpha(z) &=&    (1{-}z)^{  {1\over 2} {+} k{+}{b^2\over 2} } z^{  \frac{1}{2}{+}\alpha p_2{+}{b^2\over 2} }
   {}_{2} F_1 \left( \ft12 {+} k {+}p_0{+}\alpha p_2, \ft12 {+} k{-} p_0{+}\alpha p_2 , 1{+}2  \alpha p_2 ;  z \right)
   \\
      \widetilde{\Psi}^L_\alpha {}  (z)&=&    (z{-}1)^{  \frac{1}{2}{+}\alpha k{+}{b^2\over 2} }
 z^{  {1\over 2} {+} p_2{+}{b^2\over 2} }     {}_2 F_1 \left(  \ft12 {+} p_0{+}p_2{+}\alpha k,
 \ft12 {-} p_0{+}p_2{+}\alpha k ,1{+}2  \alpha k,1-z  \right)  \nn\\
\Psi^{L}_{\alpha}(z)   &=&  (z{-}1)^{\frac{1}{2} +  k{+}{b^2\over 2} } z^{  \alpha p_0-k {+}{b^2\over 2} }    {}_{2} F_1 \left( \ft12 {+} k {+}p_2 {-}\alpha p_0, \ft12 {+} k {-}p_2{-}\alpha p_0 , 1{-}2  \alpha p_0 ;   {1\over z} \right) \nn
  \eea
  with $\alpha=\pm$. In addition, we introduce a fourth pair of solutions
  \be
  \widetilde{\Psi}^R_\alpha {}  (z)=(-)^{ \frac{1}{2}{+}\alpha k +{b^2\over 2} }  \widetilde{\Psi}^L_\alpha {}  (z)
  \ee
  In graphical notation
  \bea
 \Psi^R_\alpha(z) &=&  \begin{tikzpicture}[baseline={(current bounding box.center)}, node distance=0.8cm and 0.8cm]
\coordinate[label=above:$k$] (k0);
\coordinate[below=of k0] (s0);
\coordinate[left=of s0] (p0);
\coordinate[right=1cm of s0] (s1);
\coordinate[above=of s1,label=above:$k_{12}$] (k1);
\coordinate[right=of s1] (p2);
\draw[line] (k0) -- (s0);
\draw[line,dashed] (s0) --node[label={[xshift=0.2cm, yshift=0.1cm]left:\scriptsize{$ 1$}}] {}  (k0);
\draw[line] (s0) -- (p0);
\draw[line,dashed] (k1) -- node[label={[xshift=0.2cm, yshift=0.1cm]left:\scriptsize{$ z$}}] {} (s1);
\draw[line] (s1) -- (p2);
\draw[line] (s0) -- node[label={[yshift=0.2cm]below:$p_2^{-\alpha}$}] {} (s1);
\draw[line] (s0) -- node[label=below:$p_0$] {} (p0);
\draw[line] (s1) -- node[label=below:$p_2$] {} (p2);
\end{tikzpicture} \qquad , \qquad  \Psi^{L}_{\alpha}(z)  =   \begin{tikzpicture}[baseline={(current bounding box.center)}, node distance=0.8cm and 0.8cm]
\coordinate[label=above:$k_{12}$] (k0);
\coordinate[below=of k0] (s0);
\coordinate[left=of s0] (p0);
\coordinate[right=1cm of s0] (s1);
\coordinate[above=of s1,label=above:$k$] (k1);
\coordinate[right=of s1] (p2);
\draw[line,dashed] (k0) -- node[label={[xshift=0.2cm, yshift=0.1cm]left:\scriptsize{$ z$}}] {} (s0);
\draw[line] (s0) -- (p0);
\draw[line] (s1) --node[label={[xshift=0.2cm, yshift=0.1cm]left:\scriptsize{$ 1$}}] {}  (k1);
\draw[line] (s1) -- (p2);
\draw[line] (s0) -- node[label={[yshift=0.1cm]below:$p_0^{\alpha}$}] {} (s1);
\draw[line] (s0) -- node[label=below:$p_0$] {} (p0);
\draw[line] (s1) -- node[label=below:$p_2$] {} (p2);
\end{tikzpicture} \nn\\
 \widetilde{\Psi}^R_{\alpha} {}  (z) &=&  \begin{tikzpicture}[scale=0.5,baseline={(current bounding box.center)}, node distance=0.6cm and 0.6cm]
\coordinate[label=left:$k$] (k0);
\coordinate[below right=of k0] (s0);
\coordinate[above right=of s0,label=right:$k_{12}$] (k1);
\coordinate[below=of s0] (s1);
\coordinate[left=of s1] (p0);
\coordinate[right=of s1,xshift=0.2cm] (s2);

\draw[line] (k0) -- node[label={[xshift=-0.2cm, yshift=0.2cm]below:\scriptsize{$ 1$}}] {} (s0);
\draw[line,dashed] (s0) -- node[label={[xshift=0.2cm, yshift=0.2cm]below:\scriptsize{$ z$}}] {} (k1);
\draw[line] (s0) -- node[label={[xshift=-0.2cm]right:\scriptsize{$ k_{\alpha}$}}] {} (s1);

\draw[line] (s1) -- node[label=below:$p_0$] {} (p0);

\draw[line] (s1) -- node[label=below:$p_2$] {} (s2);
\end{tikzpicture} \qquad , \qquad
 \widetilde{\Psi}^L_{\alpha} {}  (z) =  \begin{tikzpicture}[scale=0.5,baseline={(current bounding box.center)}, node distance=0.6cm and 0.6cm]
\coordinate[label=left:$k_{12}$] (k0);
\coordinate[below right=of k0] (s0);
\coordinate[above right=of s0,label=right:$k$] (k1);
\coordinate[below=of s0] (s1);
\coordinate[left=of s1] (p0);
\coordinate[right=of s1,xshift=0.2cm] (s2);

\draw[line,dashed] (k0) -- node[label={[xshift=-0.2cm, yshift=0.2cm]below:\scriptsize{$ z$}}] {} (s0);
\draw[line] (s0) -- node[label={[xshift=0.2cm, yshift=0.2cm]below:\scriptsize{$ 1$}}] {} (k1);
\draw[line] (s0) -- node[label={[xshift=-0.2cm]right:\scriptsize{$ k_{\alpha}$}}] {} (s1);

\draw[line] (s1) -- node[label=below:$p_0$] {} (p0);

\draw[line] (s1) -- node[label=below:$p_2$] {} (s2);
\end{tikzpicture}
  \eea
  The four pairs of solutions are related via the hypergeometric identities, so which one to use depends on our convenience.
  In particular, using the hypergeometric identities (\ref{hypid}) one finds the braiding and fusion relations
  \bea
\label{braid_relation}
\begin{tikzpicture}[baseline={(current bounding box.center)}, node distance=0.8cm and 0.8cm]
\coordinate[label=above:$k_{12}$] (k0);
\coordinate[below=of k0] (s0);
\coordinate[left=of s0] (p0);
\coordinate[right=1cm of s0] (s1);
\coordinate[above=of s1,label=above:$k$] (k1);
\coordinate[right=of s1] (p2);
\draw[line,dashed] (k0) -- node[label={[xshift=0.2cm, yshift=0.1cm]left:\scriptsize{$ z$}}] {} (s0);
\draw[line] (s0) -- (p0);
\draw[line] (s1) --node[label={[xshift=0.2cm, yshift=0.1cm]left:\scriptsize{$ 1$}}] {}  (k1);
\draw[line] (s1) -- (p2);
\draw[line] (s0) -- node[label={[yshift=0.1cm]below:$p_0^{\alpha}$}] {} (s1);
\draw[line] (s0) -- node[label=below:$p_0$] {} (p0);
\draw[line] (s1) -- node[label=below:$p_2$] {} (p2);
\end{tikzpicture} &=& \sum_{\alpha}B_{\alpha \alpha'}[_{p_0 ~p_2}^{k_{12}~k}]
\begin{tikzpicture}[baseline={(current bounding box.center)}, node distance=0.8cm and 0.8cm]
\coordinate[label=above:$k$] (k0);
\coordinate[below=of k0] (s0);
\coordinate[left=of s0] (p0);
\coordinate[right=1cm of s0] (s1);
\coordinate[above=of s1,label=above:$k_{12}$] (k1);
\coordinate[right=of s1] (p2);
\draw[line] (k0) -- (s0);
\draw[line,dashed] (s0) --node[label={[xshift=0.2cm, yshift=0.1cm]left:\scriptsize{$ 1$}}] {}  (k0);
\draw[line] (s0) -- (p0);
\draw[line,dashed] (k1) -- node[label={[xshift=0.2cm, yshift=0.1cm]left:\scriptsize{$ z$}}] {} (s1);
\draw[line] (s1) -- (p2);
\draw[line] (s0) -- node[label={[yshift=0.15cm]below:$p_2^{-\alpha}$}] {} (s1);
\draw[line] (s0) -- node[label=below:$p_0$] {} (p0);
\draw[line] (s1) -- node[label=below:$p_2$] {} (p2);
\end{tikzpicture} \nn\\
 \begin{tikzpicture}[baseline={(current bounding box.center)}, node distance=0.8cm and 0.8cm]
\coordinate[label=above:$k_{12}$] (k0);
\coordinate[below=of k0] (s0);
\coordinate[left=of s0] (p0);
\coordinate[right=1cm of s0] (s1);
\coordinate[above=of s1,label=above:$k$] (k1);
\coordinate[right=of s1] (p2);
\draw[line,dashed] (k0) -- node[label={[xshift=0.2cm, yshift=0.1cm]left:\scriptsize{$ z$}}] {} (s0);
\draw[line] (s0) -- (p0);
\draw[line] (s1) --node[label={[xshift=0.2cm, yshift=0.1cm]left:\scriptsize{$ 1$}}] {}  (k1);
\draw[line] (s1) -- (p2);
\draw[line] (s0) -- node[label={[yshift=0.1cm]below:$p_0^{\alpha}$}] {} (s1);
\draw[line] (s0) -- node[label=below:$p_0$] {} (p0);
\draw[line] (s1) -- node[label=below:$p_2$] {} (p2);
\end{tikzpicture} &=&\sum_{\alpha}F_{\alpha \alpha'}[_{p_0 ~p_2}^{k_{12}~k}]
 \begin{tikzpicture}[scale=0.5,baseline={(current bounding box.center)}, node distance=0.6cm and 0.6cm]
\coordinate[label=left:$k_{12}$] (k0);
\coordinate[below right=of k0] (s0);
\coordinate[above right=of s0,label=right:$k$] (k1);
\coordinate[below=of s0] (s1);
\coordinate[left=of s1] (p0);
\coordinate[right=of s1,xshift=0.2cm] (s2);
\draw[line,dashed] (k0) -- node[label={[xshift=-0.2cm, yshift=0.2cm]below:\scriptsize{$ z$}}] {} (s0);
\draw[line] (s0) -- node[label={[xshift=0.2cm, yshift=0.2cm]below:\scriptsize{$ 1$}}] {} (k1);
\draw[line] (s0) -- node[label={[xshift=-0.2cm]right:\scriptsize{$ k_{\alpha'}$}}] {} (s1);
\draw[line] (s1) -- node[label=below:$p_0$] {} (p0);
\draw[line] (s1) -- node[label=below:$p_2$] {} (s2);
\end{tikzpicture} \nn\\
\begin{tikzpicture}[scale=0.5,baseline={(current bounding box.center)}, node distance=0.6cm and 0.6cm]
\coordinate[label=left:$k$] (k0);
\coordinate[below right=of k0] (s0);
\coordinate[above right=of s0,label=right:$k_{12}$] (k1);
\coordinate[below=of s0] (s1);
\coordinate[left=of s1] (p0);
\coordinate[right=of s1,xshift=0.2cm] (s2);
\draw[line] (k0) -- node[label={[xshift=-0.2cm, yshift=0.2cm]below:\scriptsize{$ 1$}}] {} (s0);
\draw[line,dashed] (s0) -- node[label={[xshift=0.2cm, yshift=0.2cm]below:\scriptsize{$ z$}}] {} (k1);
\draw[line] (s0) -- node[label={[xshift=-0.2cm]right:\scriptsize{$ k_{\alpha'}$}}] {} (s1);
\draw[line] (s1) -- node[label=below:$p_0$] {} (p0);
\draw[line] (s1) -- node[label=below:$p_2$] {} (s2);
\end{tikzpicture}
 &=& \sum_{\alpha'}F^{-1}_{\alpha \alpha'}[_{p_0 ~p_2}^{k~k_{12} }]
\begin{tikzpicture}[baseline={(current bounding box.center)}, node distance=0.8cm and 0.8cm]
\coordinate[label=above:$k$] (k0);
\coordinate[below=of k0] (s0);
\coordinate[left=of s0] (p0);
\coordinate[right=1cm of s0] (s1);
\coordinate[above=of s1,label=above:$k_{12}$] (k1);
\coordinate[right=of s1] (p2);
\draw[line] (k0) -- (s0);
\draw[line,dashed] (s0) --node[label={[xshift=0.2cm, yshift=0.1cm]left:\scriptsize{$ 1$}}] {}  (k0);
\draw[line] (s0) -- (p0);
\draw[line,dashed] (k1) -- node[label={[xshift=0.2cm, yshift=0.1cm]left:\scriptsize{$ z$}}] {} (s1);
\draw[line] (s1) -- (p2);
\draw[line] (s0) -- node[label={[yshift=0.2cm]below:$p_2^{-\alpha'}$}] {} (s1);
\draw[line] (s0) -- node[label=below:$p_0$] {} (p0);
\draw[line] (s1) -- node[label=below:$p_2$] {} (p2);
\end{tikzpicture}
\eea
 with braiding and fusion matrices
 \bea
B_{\alpha \alpha'}[_{p_0 ~p_2}^{k_{12}~k}]   &=& \frac{e^{i \pi   \left( \alpha p_0- \alpha' p_2{+}{b^2\over 2}\right)}\Gamma \left(1-2  \alpha p_0\right) \Gamma \left(-2  \alpha' p_2\right)}
{\Gamma \left(\frac{1}{2} -\alpha p_0-\alpha' p_2+k \right)
	\Gamma \left(\frac{1}{2}-\alpha p_0-\alpha' p_2-k \right)}  \label{braidingB} \nn\\
 F_{\alpha \alpha'}[_{p_0 ~p_2}^{k_{12}~k}]  &=& \frac{\Gamma \left(1-2  \alpha p_0\right)
 \Gamma \left(2  \alpha' k\right)}
{\Gamma \left(\frac{1}{2}{-}\alpha p_0{+}\alpha'  k{+}p_2 \right)
	\Gamma \left(\frac{1}{2}{-}\alpha p_0{+}\alpha' k{-} p_2\right)} \nn\\
F^{-1}_{\alpha \alpha'}[_{p_0 ~p_2}^{k~k_{12} }]   &=& \frac{\Gamma (1-2  \alpha k)
\Gamma \left(-2  \alpha' p_2\right)}{\Gamma \left(\frac{1}{2}-\alpha k+p_0-\alpha' p_2
\right) \Gamma \left(\frac{1}{2}-\alpha k-p_0-\alpha' p_2\right)}
\eea
   The crucial observation \cite{Moore:1988qv,Moore:1988uz} is that the braiding and fusion moves are local operations, independent of the positions of the vertex insertions, so they are still
   valid if we extend the conformal block by adding any number of vertices to the left or right of the diagram. In this way, one can derive exact connection formulae relating the asymptotics near any two singularities (different point insertions of the degenerate vertex) for a differential equation with an arbitrary number of Fuchsian singularities !

  Confluent limits where two Fuchsian singularities collide can be studied in a similar way. The limit is obtained by taking for example $k$, $p_2$ and $z$  large,
  keeping finite the combinations $c=k+p_2$ and $\hat{z}=z/(2k)$. In this limit, the conformal blocks reduce to
  \bea
  \Psi^{L,\rm conf}_{\alpha}(\hat{z} ) &=& z^{{1\over 2}+p_0+b^2} \, e^{-{1\over 2 z} } {}_1F_1(\ft12+c-\alpha p_0,1-2\alpha p_0,\ft{1}{\hat z})\nn\\
   \Psi^{R,\rm conf}_{\alpha}(\hat{z} ) &=& z^{1+\alpha c+b^2} \, e^{-{\alpha\over 2 z} } {}_2F_0(\ft12+\alpha c+ p_0,\ft12+\alpha c- p_0, {-}\alpha \hat{z})
   \label{allsolutionsc}
  \eea
that are related to each other by the braiding relation
  \be\label{braiding2}
  \Psi^{L,\rm conf}_{\alpha}(\hat{z} ) =  \sum_{\alpha'} B^{\rm conf}_{\alpha \alpha'}[_{p_0 ~c}^{k_{12}}]    \Psi^{R,\rm conf}_{\alpha'}(\hat{z} )
\ee
    with
 \bea
B^{\rm conf}_{\alpha' \alpha''}[_{p_0 ~c}^{k_{12}}]   &=&e^{-{\rm i} \pi \delta_{1\alpha''}({1\over 2} {+} c{-} \alpha' p_0)}  \frac{   \Gamma \left(1-2  \alpha' p_0\right) }
{\Gamma \left(\frac{1}{2} -\alpha' p_0-\alpha'' c \right) }  \label{braidingBconf}
\eea
 We remark that the second line of (\ref{allsolutionsc})   is understood as a series expansion around $\hat{z}\approx 0$, since
 the $ {}_{2} F_0$ series does not converge.  In this paper we focus on the NS limit (\ref{NSlimit}), so we can discard the $b^2$-terms in all formulas above and the $\partial_z$-term in
 (\ref{BPZeq4}).

\subsection{Five-point conformal block}
\label{sec5}

 Next we consider the  $n=2$ case (degenerate five-point conformal block) leading to a differential equation with four Fuchsian singularities (Heun equation).
 The connection matrices in this case have been obtained in section 3.1 of \cite{Bonelli:2021uvf}. They have been also related to the Q-matrix
 of integrable systems  in \cite{Fioravanti:2020udo,Fioravanti:2021bzq,Fioravanti:2021dce}.

Taking the singularities at
\be
z_0=\infty \quad , \quad z_1=1 \quad , \quad z_2=z\quad , \quad z_3 =x \quad , \quad z_4=0
\ee
 the Ward identities (\ref{conformal_Ward_identities}) can be solved as
 \be
 \partial_{z_0 }  =0 \quad , \quad   \partial_{z_1} =\Delta {-} z \partial_{z} {-} x \partial_{x}    \quad , \quad
 \partial_{z_4}  = (z{-}1) \partial_{z}+(x{-}1) \partial_{x}  -\Delta
 \ee
 with
 \be
 \Delta=   \Delta_{p_0}- \Delta_{p_3}-\Delta_{k_0} -\Delta_{k}-\Delta_{12}
 \ee
 Plugging this into (\ref{BPZ}), taking the limit (\ref{NSlimit}) and using (\ref{f0f1}), the BPZ equation takes the  normal form (\ref{can}) with
 \bea
\label{Q_function_cft2}
Q(z,x)=\frac{\delta_{p_3} }{z^2}+\frac{\delta_{k_0} }{(z-1)^2}
+\frac{\delta_k}{(z-x)^2}
+\frac{\delta }{z(z-1)}+\frac{(1-x) U}{z(z-1)(z-x)}
\eea
and
\be
U(x) =\lim_{b \to 0}\,b^2 \,x \partial_{x}\ln \Psi(z,x)  \qquad , \qquad \delta = \delta_{p_0} -  \delta_{p_3}-\delta_{k_0} -\delta_{k} \label{umatone}
\ee
 The first formula in (\ref{umatone}) is nothing but the quantum version of the Matone relation \cite{Matone:1995rx,Flume:2004rp}.

 Again there are several basis of solutions one can use depending on the position of the degenerated field insertion. We will mainly use the following
 three representations \cite{Bonelli:2021uvf}
 \bea
  \Psi_\alpha(z,x) &=&   \begin{tikzpicture}[scale=0.5,baseline={(current bounding box.center)}, node distance=0.6cm and 0.6cm]
\coordinate[label=left:$k_0$] (k0);
\coordinate[below right=of k0] (s0);
\coordinate[above right=of s0,label=right:$k_{12}$] (k1);
\coordinate[below=of s0] (s1);
\coordinate[left=of s1] (p0);
\coordinate[right=of k1,xshift=0.2cm,label=right:$k$] (k2);
\coordinate[right=of s1,xshift=0.8cm] (s2);
\draw[line] (k2) -- node[label={[xshift=0.2cm, yshift=0.2cm]below:\scriptsize{$ x$}}] {} (s2);

\draw[line] (k0) -- node[label={[xshift=-0.2cm, yshift=0.2cm]below:\scriptsize{$ 1$}}] {} (s0);
\draw[line,dashed] (s0) -- node[label={[xshift=0.2cm, yshift=0.2cm]below:\scriptsize{$ z$}}] {} (k1);
\draw[line] (s0) -- node[label={[xshift=-0.2cm]right:\scriptsize{$ k_{0\alpha}$}}] {} (s1);

\draw[line] (s1) -- node[label=below:$p_0$] {} (p0);

\draw[line] (s1) -- node[label=below:$p_2$] {} (s2);
\coordinate[right=of s2] (p3);
\draw[line] (s2) -- node[label=below:$p_3$] {} (p3);
\end{tikzpicture}
=
  \sum_{\alpha'}F^{-1}_{\alpha \alpha'}[_{p_0 ~p_2}^{ k_0 ~k_{12} }]
   \begin{tikzpicture}[baseline={(current bounding box.center)}, node distance=0.8cm and 0.8cm]
\coordinate[label=above:$k_0$] (k0);
\coordinate[below=of k0] (s0);
\coordinate[left=of s0] (p0);
\coordinate[right=1cm of s0] (s1);
\coordinate[above=of s1,label=above:$k_{12}$] (k1);
\coordinate[right=of s1] (p2);
\coordinate[right=1cm of s1] (s2);
\coordinate[above=of s2,label=above:$k$] (k2);
\coordinate[right=of s2] (p3);
\draw[line] (k0) -- (s0);
\draw[line,dashed] (s0) --node[label={[xshift=0.2cm, yshift=0.1cm]left:\scriptsize{$ 1$}}] {}  (k0);
\draw[line] (s0) -- (p0);
\draw[line,dashed] (k1) -- node[label={[xshift=0.2cm, yshift=0.1cm]left:\scriptsize{$ z$}}] {} (s1);
\draw[line] (s1) -- (p2);
\draw[line] (s0) -- node[label={[yshift=0.2cm]below:$p_2^{-\alpha'}$}] {} (s1);
\draw[line] (s0) -- node[label=below:$p_0$] {} (p0);
\draw[line] (k2) -- node[label={[xshift=0.2cm, yshift=0.1cm]left:\scriptsize{$ x$}}] {} (s2);
\draw[line] (s2) -- (p3);
\draw[line] (s1) -- node[label=below:$p_2$] {} (s2);
\draw[line] (s2) -- node[label=below:$p_3$] {} (p3);
\end{tikzpicture} \nn\\
    &=&   \sum_{\alpha',\alpha''}F^{-1}_{\alpha  \alpha'}[_{p_0 ~p_2}^{ k_0 ~k_{12} }]
B_{\alpha'  \alpha''}[_{p_2^{-\alpha'} p_3}^{k_{12}~~k}]
   \begin{tikzpicture}[baseline={(current bounding box.center)}, node distance=0.8cm and 0.8cm]
\coordinate[label=above:$k_0$] (k0);
\coordinate[below=of k0] (s0);
\coordinate[left=of s0] (p0);
\coordinate[right=1cm of s0] (s1);
\coordinate[above=of s1,label=above:$k$] (k1);
\coordinate[right=of s1] (p2);
\coordinate[right=1cm of s1] (s2);
\coordinate[above=of s2,label=above:$k_{12}$] (k2);
\coordinate[right=of s2] (p3);
\draw[line] (k0) -- (s0);
\draw[line,dashed] (s0) --node[label={[xshift=0.2cm, yshift=0.1cm]left:\scriptsize{$ 1$}}] {}  (k0);
\draw[line] (s0) -- (p0);
\draw[line] (k1) -- node[label={[xshift=0.2cm, yshift=0.1cm]left:\scriptsize{$ x$}}] {} (s1);
\draw[line] (s1) -- (p2);
\draw[line] (s0) -- node[label={[yshift=0.2cm]below:$p_2^{-\alpha'}$}] {} (s1);
\draw[line] (s0) -- node[label=below:$p_0$] {} (p0);
\draw[line,dashed] (k2) -- node[label={[xshift=0.2cm, yshift=0.1cm]left:\scriptsize{$ z$}}] {} (s2);
\draw[line] (s2) -- (p3);
\draw[line] (s1) -- node[label={[yshift=0.2cm]below:$p_3^{-\alpha''}$}] {} (s2);
\draw[line] (s2) -- node[label=below:$p_3$] {} (p3); \label{fusb2}
\end{tikzpicture} \label{connection}
   \eea
       To compute the asymptotics we use the degenerate three point  correlator
   \be
     \begin{tikzpicture}[scale=0.5,baseline={(current bounding box.center)}, node distance=0.6cm and 0.6cm]
\coordinate[label=left:$P$] (k0);
\coordinate[below right=of k0] (s0);
\coordinate[above right=of s0,label=right:$k_{12}$] (k1);
\coordinate[below=of s0] (s1);
\draw[line] (k0) -- node[label={[xshift=-0.2cm, yshift=0.2cm]below:\scriptsize{$ z_s$}}] {} (s0);
\draw[line,dashed] (s0) -- node[label={[xshift=0.2cm, yshift=0.2cm]below:\scriptsize{$ z$}}] {} (k1);
\draw[line] (s0) -- node[label={[xshift=-0.2cm]right:\scriptsize{$ P^{\alpha}$}}] {} (s1);

\end{tikzpicture} \approx (z-z_s)^{\Delta_{P_\alpha}-\Delta_P-\Delta_{12} } \approx    (z-z_s)^{{1\over 2} -\alpha P  }  \label{3point}
\ee
 Here and below we will always discard $b^2$-terms  since we are interested in the $b\to 0$ limit.
   Plugging (\ref{3point}) into the left and right hand sides of (\ref{connection}) one finds the asymptotics of the solutions at the two boundaries
       \bea
  \Psi_\alpha(z,x)  &   \underset{z\to 1}{\approx}  &   \, f_\alpha(z)   \,
    \begin{tikzpicture}[baseline={(current bounding box.center)}, node distance=0.8cm and 0.8cm]
\coordinate[label=above:$k_{0\alpha}$] (k0);
\coordinate[below=of k0] (s0);
\coordinate[left=of s0] (p0);
\coordinate[right=1cm of s0] (s1);
\coordinate[above=of s1,label=above:$k$] (k1);
\coordinate[right=of s1] (p2);
\draw[line] (k0) -- node[label={[xshift=0.2cm, yshift=0.1cm]left:\scriptsize{$ 1$}} ] {} (s0);
\draw[line] (s0) -- (p0);
\draw[line] (s1) --node[label={[xshift=0.2cm, yshift=0.1cm]left:\scriptsize{$ x$}}] {}  (k1);
\draw[line] (s1) -- (p2);
\draw[line] (s0) -- node[label=below:$p_2$] {} (s1);
\draw[line] (s0) -- node[label=below:$p_0$] {} (p0);
\draw[line] (s1) -- node[label=below:$p_3$] {} (p2);
\end{tikzpicture}  \nn\\
 \Psi_\alpha(z,x)   & \underset{z\to 0}{\approx} & \sum_{\alpha',\alpha''} \tilde{f}_{\alpha''} (z) \, F^{-1}_{\alpha  \alpha'}[_{p_0 ~p_2}^{ k_0 ~k_{12} }]
B_{\alpha'  \alpha''}[_{p_2^{-\alpha'} p_3}^{k_{12}~~k}]
    \,  \begin{tikzpicture}[baseline={(current bounding box.center)}, node distance=0.8cm and 0.8cm]
\coordinate[label=above:$k_0$] (k0);
\coordinate[below=of k0] (s0);
\coordinate[left=of s0] (p0);
\coordinate[right=1cm of s0] (s1);
\coordinate[above=of s1,label=above:$k$] (k1);
\coordinate[right=of s1] (p2);
\coordinate[right=1cm of s1] (s2);
\coordinate[right=of s2] (p3);
\draw[line] (k0) -- (s0);
\draw[line,dashed] (s0) --node[label={[xshift=0.2cm, yshift=0.1cm]left:\scriptsize{$ 1$}}] {}  (k0);
\draw[line] (s0) -- (p0);
\draw[line] (k1) -- node[label={[xshift=0.2cm, yshift=0.1cm]left:\scriptsize{$ x$}}] {} (s1);
\draw[line] (s1) -- (p2);
\draw[line] (s0) -- node[label={[yshift=0.2cm]below:$p_2^{-\alpha'}$}] {} (s1);
\draw[line] (s0) -- node[label=below:$p_0$] {} (p0);
\draw[line] (s1) -- node[label={[yshift=0.2cm]below:$p_3^{-\alpha''}$}] {} (s2);
 \end{tikzpicture}
 \label{asym}
   \eea
   with
 \be
 f_\alpha(z) = (1-z)^{ {\frac{1}{2} {-}\alpha  k_0}  }  \quad ,\quad   \tilde f_{\alpha''}(z)=z^{\frac{1}{2} {+}\alpha'' p_3 } \label{asym01}
 \ee
 The conformal blocks appearing in right hand sides of (\ref{connection})  can be written as instanton sums. For example
\bea
 &&  {\cal F} {}_{p_0}{}^{k_0} {}_{p_2^{-\alpha'} } {}^{k_{12} } {}_{p_2}{}^{k}{}_{p_3}  (1,z,x)
 =   x^{\Delta_{p_2} {-}\Delta_{p_3}{-}\Delta_{12}} \, (1{-}x)^{ (2k_0{-}bQ)(2k_2{+}bQ)\over 2 b^2} \\
 &&\qquad \times\,  z^{ {1\over 2} + \alpha'  p_2+ {b^2 \over 2}  }\,  \,   (1{-}z)^{(2 k_0  -b Q)(1+b^2) \over b^2 }\left(1{-}{x\over z} \right)^{{1\over 2}+  k+{b^2\over 2}   } Z_{\rm inst} {}_{p_0}{}^{k_0} {}_{p_2^{-\alpha'} } {}^{k_{12} } {}_{p_2}{}^{k}{}_{p_3} (1,z,x) \nn
\eea
  In applications to gravity, the terms in the first line can be discarded since they do not depend on $z$.
  The instanton part can be computed from (\ref{zinstG}) order by order in $q_1=z$, $q_2=x/z$.  Up to one instanton order one finds
\bea
&& \ln  {\cal F}_{p_0}{}^{k_0} {}_{p_2^{-\alpha'} } {}^{k_{12} } {}_{p_2}{}^{k}{}_{p_3}  (1,z,x)    \approx
    {x \rho_1 \rho_2 \over 8 b^2( 1-4 p_2^2) }  +   \left( {p_3^2 {-} p_2^2 \over b^2} {-} 1\right) \log x\label{zinstexplicit}
 \\
 &&{+} \left(\ft{1}{2} {+} \alpha'  p_2 \right)\log z  {-}
{z  \rho_1 \over 4(2 \alpha' p_2+1) } +  {x  \rho_2 \over 4 z (2\alpha' p_2-1) } - x {4 \rho_1\alpha' p_2+8 \rho_2 p_2^2 +\rho_1 \rho_2 -2 \rho_2-2\rho_1 \over 8 (1-2\alpha' p_2) (1-4 p_2^2) }+\ldots \nn
\eea
with
\be
\rho_1=1+4 p_0^2-4 k_0^2 -4 p_2^2 \qquad , \qquad  \rho_2=1+4 p_3^2-4 k^2 -4 p_2^2
\ee
 We notice that the terms in the first line of (\ref{zinstexplicit}) grow as $b^{-2}$ in the NS limit and depend only on $x$ consistently with (\ref{f0f1}).
  For applications to gravity, we will mainly be interested on the limit $x, z \to 0$ keeping $\frac{x}{z}$ finite. In this limit, an explicit evaluation of the first few instanton corrections shows that $Z_{\rm inst}$ can be resummed into a hypergeometric function leading to
\be \label{entire}
 {\cal F} {}_{p_0}{}^{k_0} {}_{p_2^{-\alpha'} } {}^{k_{12} } {}_{p_2}{}^{k}{}_{p_3}
  \underset{z\to 0}{\sim}     z^{{1\over 2} {+} \alpha'  p_2 }  \left(1{-}{x\over z} \right)^{{1\over 2}+  k  }
 {}_2 F_1 (  \ft{1}{2}{+} k{+}p_3{-}\alpha' p_2,
  \ft{1}{2}{+}k{-}p_3{-}\alpha' p_2,  1{-}2\alpha' p_2, \ft{x}{z} )
\ee
    As before, one can consider  the confluent limit of the five-point conformal block. This limit is obtained by taking  $k$, $p_3$  and $z$ large but $c=k+p_3$  and $\hat{z} =z/(2k)$ finite. Combining (\ref{connection}) and (\ref{entire}) one finds
  \be
\Psi_\alpha (\hat{z} )   \underset{ \hat z\to 0}{\approx}   \sum_{\alpha'}  \hat z^{{1\over 2} + \alpha'  p_2 }  e^{{-}{x\over 2\hat{z} } }  F^{-1}_{\alpha  \alpha'}[_{p_0 ~p_2}^{ k_0 ~k_{12} }]
   {}_1 F_1 (  \ft{1}{2}{+} c {-}\alpha' p_2,
    1{-}2\alpha'  p_2, \ft{ x}{ \hat z} )  \label{psiminus00}
\ee
On the other hand in the limit where $\hat{z} << x$ using the asymptotic expansion of the hypergeometric function one finds
     \bea\label{allsolutions2}
\Psi_\alpha (\hat{z} )  & \underset{ \hat z\to 0}{\approx}  & \sum_{\alpha',\alpha''}
 x^{{1\over 2} + \alpha'  p_2 }  \left(\ft{\hat {z} }{x} \right)^{1+ \alpha'' c }  e^{{-}\alpha''{ x\over 2\hat{z} } }  F^{-1}_{\alpha  \alpha'}[_{p_0 ~p_2}^{ k_0 ~k_{12} }] B^{\rm conf}_{\alpha' \alpha''}[_{p_2 ~c}^{k_{12}}] \nn\\
 && \qquad \qquad \qquad  \times
 {}_{2} F_0 \left( \ft12 {+}\alpha'' c {+}p_0, \ft12 {+} \alpha'' c{-} p_0 ,  {-} \alpha'' \ft{\hat {z} }{x} \right)
\eea
    with
 \bea
B^{\rm conf}_{\alpha' \alpha''}[_{p_2 ~c}^{k_{12}}]   &=& e^{-{\rm i} \pi \delta_{1\alpha''}({1\over 2} {+}  c{-} \alpha' p_2)}  \frac{  \Gamma \left(1-2  \alpha' p_2\right) }
{\Gamma \left(\frac{1}{2} -\alpha' p_2-\alpha'' c \right) }  \label{braidingBconf}
\eea

\subsection{BPZ equation from the quantum deformed SW curve }

 The BPZ equation  can alternatively be obtained as a quantum deformation of the geometry describing the $U(2)$ gauge theory with four fundamental hypermultiplets resulting from omitting the node associated to the "degenerate" insertion, let us say node 1.
  In the limit $\epsilon_1 \to 0$ and $\epsilon_2=\hbar$  the dynamics of this theory is described by the difference equation \cite{Poghossian:2010pn,Fucito:2011pn}
 \be
 y(v) y(v-\hbar)-P_2(v) y(v-\hbar) +q P_0(v-\hbar)  P_3(v) =0  \label{difference}
 \ee
 with
\bea
P_0(v)&=& (v-a_{01}) (v-a_{02}) \qquad, \qquad P_3(v)= (v-a_{31}) (v-a_{32}) \,;\\
P_2(v)&=& v^2(1{+}q){-}q v   (a_{01}{+}a_{02}{+}a_{31}{+}a_{32}) {-}u {+}q \left[u
{+}(a_{01}{+}a_{02})(a_{31}{+}a_{32} {-}\hbar)
{+}a_{01}a_{02} {+} a_{31}a_{32}
\right]\nn
\eea
 Here $a_{0u}$ and $a_{3u}$
  play the role of masses while $u$ parametrizes the quantum Coulomb branch. $y(v)$ is the generating function of the expectation values
$\langle \tr \phi^J\rangle $ of the scalar $\phi$ in the vector multiplet
\bea
\frac{d\log y(v)}{dv}=\sum_{J=0}
^\infty \frac{\langle \tr \phi^J\rangle}{v^{J+1}} = {2\over v} +{2u \over v^3} +\ldots \label{logy}
\eea
with
\be
\langle\tr  1 \rangle =2 \quad , \quad  \langle\tr \phi \rangle =0  \quad, \quad \langle\tr \phi^2 \rangle =2  u
\ee
The form of $P_2(v)$ is determined by requiring that the expansion (\ref{logy}) satisfies the difference equation (\ref{difference}) at large $v$.
In the $\hbar\to 0$ limit (\ref{difference}) reduces to the SW curve. This is why
the case in which $\hbar$ is kept finite is usually referred as the deformed (or ``quantum") geometry.
As in the undeformed case, the function $y(v)$ determines the differential
\bea
\lambda=v\,d \log y(v)\,.
\eea
Solving (\ref{difference}) order by order in the limit of small $q$, one finds that
at any instanton order the number of poles of $\lambda$ is finite and is of
the form $v_r = \pm \sqrt{u}+r \hbar $ with $r=0,1,\ldots ,k$ and $k$ being the instanton order.
The expectation value is given by
\bea
\label{acycleint}
a=\oint_{\gamma}  {dv\over 2 \pi i} v \partial_v \log  y(v)
\eea
where $\gamma$ is the contour encircling  all the poles around $\sqrt{u}+r \hbar $
with $r=0,1,\ldots k$.
Then to find the gauge prepotential as a power series in $q$ one can first invert (\ref{acycleint}) to express
$u$ as a function of $a$ and make use of the  relation \cite{Flume:2004rp}
\be
{\cal F}^{U(2)} =-\int^q {u(a,q')\over q'} dq'
\ee
There is an alternative, more direct way to find the function $u(a)$  (again based on
(\ref{difference})) by solving the equation \cite{Poghosyan:2020zzg}
\bea
\label{fractionequality}
\frac{ q M(a+\hbar)}{P_2(a+\hbar)-\frac{ q M(a+2\hbar)}{P_2(a+2\hbar)-\ldots
}}
+\frac{ q M(a)}{P_2(a-\hbar)-\frac{ q M(a-\hbar)}{P_2(a-2\hbar)-\ldots 
}}=P_2(a)
\eea
where, for the sake of simplicity, we used the notation
\be
M(a)=P_0(a-\hbar)P_3(a)
\ee
Plugging $u$ inside the $P_2$ function, one can solve (\ref{fractionequality})  for $u(a)$ or $a(u)$ as a power series in $q$
\bea
u &=& a^2 + c_1(a) q+c_2(a) q^2+\ldots \nn\\
a &=& \sqrt{u} + \tilde{c}_1(u) q+\tilde{c}_2(u) q^2+\ldots \label{uqsmall}
\eea
Equation (\ref{fractionequality})
is a very convenient tool for the numerical evaluation of $u(a)$ for large values of $q$, where the series expansion is not useful.
It is convenient to represent the $y$-function as the ratio
  \be
  y(v) ={ P_0(v)  Y(v) \over Y(v-\hbar)}
  \ee
   and denote the Fourier transform of $Y(v)$ as
\be
F(z)=\sum_{v \in a_{1 u} +\hbar \mathbb{Z} } Y(v) z^{-{v \over \hbar} }
\ee
   In these terms, the difference equation (\ref{difference})   translates into the differential equation   \cite{Poghossian:2016rzb}
    \be
   \left[ P_0\left(-\hbar z {d\over dz} \right) -P_2\left(-\hbar z {d\over dz} \right) z^{-1} + q \,P_3\left(-\hbar z {d\over dz} \right) z^{-2} \right] F(z)   =0
 \ee
 This equation can be put into the normal form by writing
\bea
\label{psi_vs_f}
F(z)=z^{\frac{3}{2}-\frac{a_{21}+a_{22}}{2\hbar }} (1-z)^{-1-\frac{a_{01}+a_{02} }{2\hbar }} (z-q)^{-1+\frac{a_{21}+a_{22}}{2\hbar } }\Psi(z)
\eea
 Setting $\hbar=1$ one finds (\ref{Q_function_cft2}) with $x=q$ and
\bea
\label{c_delta_map}
\delta_{p_3} &=&\ft14 {-}  p_3^2= {1\over 4} {-}\left( {a_{31}{-}a_{32}  \over 2  } \right)^2  \quad,\quad
  \delta_{p_0} = \ft14{ -}  p_0^2={1\over 4}{ -}\left( {a_{01} {-}a_{02}  \over 2} \right)^2  \nn\\
 \delta_{k_0} &=&\ft14 {-} k_0^2 ={1\over 4} {-}\left( { 1  {+}a_{01}{+}a_{02}  \over 2 \ } \right)^2 \quad , \quad
   \delta_k = \ft14 {-} k^2={1\over 4} {-}\left( { 1 {-}a_{31}{-}a_{32}  \over 2  } \right)^2  \\
 U&=& u{-}\frac{1}{4}{+}\delta_k{+}\delta_{p_3}  {+}{x(a_{01}{+}a_{02})(2{-}a_{31}{-}a_{32}) \over 2   (1{-}x)}
\approx  u{-}\frac{1}{4} {+}\delta_k{+}\delta_{p_3} {-}{x( 1{-} 2 k_0  )(1{+}2 k) \over 2(1{-}x)} \nn
\eea
 where in the right hand side of each equation we used the AGT dictionary (\ref{agtdic})  and set $k_1=k_{12}\approx \ft{1}{2}$. This leads to the identification
 \be
 \Psi (z)=  \mathcal{F}_{p_0}{}^{k_0} {}_{p_2^{-\alpha} }{}^{k_{12}}  {}_{p_2}{}^{k}  {}_{p_3}(1,z,x) =  \begin{tikzpicture}[baseline={(current bounding box.center)}, node distance=0.8cm and 0.8cm]
\coordinate[label=above:$k_0$] (k0);
\coordinate[below=of k0] (s0);
\coordinate[left=of s0] (p0);
\coordinate[right=1cm of s0] (s1);
\coordinate[above=of s1,label=above:$k_{12}$] (k1);
\coordinate[right=of s1] (p2);
\coordinate[right=1cm of s1] (s2);
\coordinate[above=of s2,label=above:$k$] (k2);
\coordinate[right=of s2] (p3);
\draw[line] (k0) -- (s0);
\draw[line,dashed] (s0) --node[label={[xshift=0.2cm, yshift=0.1cm]left:\scriptsize{$ 1$}}] {}  (k0);
\draw[line] (s0) -- (p0);
\draw[line,dashed] (k1) -- node[label={[xshift=0.2cm, yshift=0.1cm]left:\scriptsize{$ z$}}] {} (s1);
\draw[line] (s1) -- (p2);
\draw[line] (s0) -- node[label={[yshift=0.2cm]below:$p_2^{-\alpha'}$}] {} (s1);
\draw[line] (s0) -- node[label=below:$p_0$] {} (p0);
\draw[line] (k2) -- node[label={[xshift=0.2cm, yshift=0.1cm]left:\scriptsize{$ x$}}] {} (s2);
\draw[line] (s2) -- (p3);
\draw[line] (s1) -- node[label=below:$p_2$] {} (s2);
\draw[line] (s2) -- node[label=below:$p_3$] {} (p3);
\end{tikzpicture}
 \ee
\subsection{Irregular singularities}

  Theories with a smaller number of hypermultiplets can be obtained from the generic $N_f=(2,2)$ case by decoupling i.e. by sending some of the ``masses'' $a_{0u}$ and/or $a_{3u}$ to infinity keeping the products  $q a_{0u}$ or   $q a_{3u}$ finite. In addition, when sending $a_{0u}$ to infinity one should also rescale $z \to z/a_{0u}$ and $Q\to Q/a_{0u}^2$.  The general case is labeled by the two-value set $\{ \nu_f \}$, $\nu_f=0,1$, $f=1,\ldots 4$, that determines which flavours are decoupled.  The $Q$-function is obtained by taking $a_{0u}$ and/or $a_{3u}$ large depending on whether $\nu_f=0,1$ and keeping the finite terms under the rescaling
  \be
  Q_{2{-}\nu_1{-}\nu_2,2{-}\nu_3{-}\nu_4}={ Q_{22} \over a_{01}^{2 \nu_1} a_{02}^{2 \nu_2} } \quad ,\quad z \to {z\over a_{01}^{\nu_1} a_{02}^{ \nu_2} } \quad, \quad  x \to {x\over a_{01}^{ \nu_1} a_{02}^{ \nu_2} a_{31}^{\nu_3} a_{32}^{ \nu_4} }
  \ee
   The results are given by
\bea
\label{Qijns}
Q_{2,2}&=&\frac{\delta_{p_3} }{z^2}+\frac{\delta_{k_0} }{(z-1)^2}
+\frac{\delta_{k} }{(z-x)^2}
+\frac{\delta }{z(z-1)}+\frac{(1-x) U}{z(z-1)(z-x)}   \qquad~~~~~~~~~~ {\rm HE} \\
Q_{2,1} &=&-\frac{x^2}{4
z^4} + \frac{x\, c }{z^3}+\frac{u-\frac{1}{4} +x\left(\ft12 -k_0 \right)}
{(z-1) z^2}+\frac{\delta_{k_0} }{(z-1)^2 z}+\frac{\delta_{p_0}}{(z-1) z}   \quad\quad ~~~~~~~{\rm CHE}  \nn\\
Q_{1,2} &=& \frac{\delta _{k} }{(x-z)^2}+\frac{\delta_{p_3}}{z^2} -
\frac{c' }{z} +\frac{u-\frac{1}{4}+\delta _{k}+\delta _{p_3}
+ x\left( k+\ft12 \right)}{z (x-z)} -\frac{1}{4}    \quad\quad  \quad ~~~~~~{\rm CHE} \nn\\
Q_{2,0} &=&  -\frac{x }{z^3}+\frac{u-\frac{1}{4}}{(z-1) z^2}
+\frac{\delta_{k_0} }{(z-1)^2 z}+\frac{\delta_{p_0}}{(z-1) z}  \qquad\qquad \qquad~ ~~~~~~~~~~~~~~~{\rm RCHE}  \nn\\
Q_{0,2} &=& \frac{\delta _{k} }{(x-z)^2}+\frac{\delta_{p_3}}{z^2}-
\frac{1}{z} +\frac{ u-\frac{1}{4} +\delta _{k} +\delta _{p_3}}{z (x-z)}  \qquad\qquad \qquad ~~~~~~~~~~~~~~~~~~{\rm RCHE}  \nn\\
Q_{1,1} &=&-\frac{x^2}{4 z^4}+ \frac{x\, c}{z^3}
+\frac{\frac{1}{4}-u+\ft{x}{2} }{z^2} +\frac{ c'}{z} -\frac{1}{4}\qquad\qquad \qquad ~~~~~~~~~~~~~~~~~~~~~~~ {\rm DCHE} \nn\\
Q_{1,0} &=&-\frac{x}{z^3} +\frac{\frac{1}{4}-u}{z^2}+  \frac{c'}{z}-\frac{1}{4}\qquad\qquad \qquad\qquad\qquad ~~~~~~~~~~~~~~~~~~~\qquad{\rm DRCHE} \nn\\
Q_{0,1} &=& -\frac{x^2}{4z^4}+\frac{x \, c}{z^3}+\frac{\frac{1}{4}-u}{z^2}-\frac{1}{z}\qquad\qquad \qquad\qquad\qquad ~~~~~~~~~~~~~~~~\qquad {\rm DRCHE} \nn\\
Q_{0,0} &=&-\frac{x}{z^3} +\frac{\frac{1}{4}-u}{z^2}-\frac{1}{z}\qquad\qquad \qquad \qquad\qquad\qquad\qquad~~~~~~~~~~~~~~~~~~~{\rm DRDCHE}\nn
\eea
 with $U$ given by (\ref{c_delta_map}) and
 \be
 c=k+p_3\qquad , \qquad  c'=p_0-k_0
 \label{cdef}\ee
 The differential equations with these $Q$ function are known in the mathematical literature as Heun Equation (HE), Confluent Heun Equation (CHE), Reduced Confluent Heun Equation (RCHE),
Double Confluent Heun Equation (DCHE), Double Reduced Confluent Heun Equation (DRCHE) and Double Reduced Double Confluent Heun Equation (DRDCHE)
and have been recently studied with techniques similar to the ones employed here in \cite{Bonelli:2022ten}.

 \section{ Wave scattering: QNM's, cross sections and echoes }

     The general solution of the BPZ equation can be written as a linear combination of $\Psi_\alpha(z)$ given in (\ref{connection}).
  In this section we consider the scattering from $z=0$ towards an inner boundary at $z=1$ characterised by some reflectivity properties.
   Near the two boundaries, we write
  \be
    \Psi(z)
  \underset{z\to 1}{\approx}  \sum_\alpha c_\alpha  f_\alpha(z) \qquad , \qquad   \Psi(z)   \underset{z\to 0}{\approx}
  \sum_{\alpha''} \tilde c_{\alpha''} \tilde f_{\alpha''}(z)
    \ee
   with  $f_\alpha(z)$ and $\tilde f_{\alpha''}(z)$,  the local solutions (\ref{asym01})
   and
\be
    \tilde c_{\alpha''} =  \sum_{\alpha}  c_\alpha M_{\alpha \alpha''}
  \ee
   The  connection matrix $M_{\alpha \alpha''}$ follows from (\ref{asym}) and reads \cite{Bonelli:2021uvf}
    \be
  M_{\alpha \alpha''} (x) =
 \sum_{\alpha'}F^{-1}_{\alpha \alpha'}[_{p_0 ~p_2}^{k_0~k_{12} }]
B_{\alpha' \alpha''}[_{p_2~  p_3}^{k_{12}~k}]
{  \cF_{p_0}{}^{k_{0} }{}_{p_2^{-\alpha'} }{}^{k}{}_{p_3^{-\alpha''} }   (1 ,x)  \over  \cF_{p_0}{}^{k_0^\alpha }{}_{p_2}{}^{k}{}_{p_3}   (1 ,x)   }
     \ee
   The reflectivity  at the inner boundary is specified  by the ratio
 \be
 \mathfrak{R}={c_+\over c_-}
 \ee
  The physical properties of the membrane determine the dependence of $\mathfrak{R}$ from the frequency, $\omega$. A typical choice is $\mathfrak{R} \sim e^{-{\rm i} \omega x_0}$ with $x_0$ specifying the size of the ECO interior. For simplicity we locate the membrane exactly at the position of the singularity ( the would be horizon).

  We will mainly consider the two extreme cases: perfectly absorbing ($c_+=0$) and perfectly reflecting ($c_-=0$) boundary conditions.
The perfectly absorbing boundary conditions represent a BH horizon from which nothing can escape.  The opposite case, represents a mirror like  ECO able to reflect all incoming light.
Motivated by this physical distinction we will refer to the two extreme choices of boundary conditions at $z=1$ as
 \bea
 {\rm Black~ hole: } \quad & c_+ = 0  \nn\\
  {\rm Perfectly ~reflecting~ ECO: } \quad & c_- = 0
 \eea

\subsection{Absorption cross section, QNM's and superradiance}

We first consider the scattering of a scalar wave in a geometry described by the BPZ wave equation. The asymptotic forms of the BPZ solutions near the two boundaries are given by (\ref{asym01}).
Alternatively one can introduce a tortoise like coordinate
 \be
 z_*=\ln(1-z)-\ln(z)
 \ee
that maps the $z$-interval boundaries 0 and 1 to $\infty$ and $-\infty$ respectively.
 In these variables, the asymptotic behaviour at the two boundaries are planar waves
 \be
{ f_\alpha(z) \over \sqrt{z(1-z) } }  \underset{z\to 1}{\sim} e^{-\alpha k_0  z_*} \qquad , \qquad   { \tilde f_{\alpha''}(z) \over \sqrt{z(1-z) } }  \underset{z\to 0}{\sim} e^{-\alpha'' p_3  z_*}
 \ee
 We will consequently refer to the  $-$  and  $+$ components  for the indices $\alpha,\alpha''$ as ingoing and outgoing waves respectively.
Using the equation of motion (\ref{can}), it is easy to show that
  the flux $\mathfrak{f} $ carried by the wave is constant. Using the asymptotic forms (\ref{asym01}) near the two boundaries  $z\approx 0,1$ and assuming $k_0,p_3$ to be purely imaginary,  one finds
\be
\mathfrak f={\rm Im} \left[ \Psi^*(z) \Psi'(z)  \right]= {\rm i}  k_0 \left( |c_+|^2-  |c_-|^2 \right) = {\rm i} p_3 \left( |\tilde c_+|^2-  |\tilde c_-|^2 \right) \label{flux}
\ee
 We refer to the contributions
 \be
  \mathfrak{f}_{\rm in} = -{\rm i} p_3 |\tilde c_-|^2 \quad , \quad  \mathfrak{f}_{\rm out} = -{\rm i} p_3 |\tilde c_+|^2 \quad , \quad
  \mathfrak{f}_{\rm abs}  =-{\rm i} k_0 | c_-|^2 \quad , \quad   \mathfrak{f}_{\rm ref}  = -{\rm i} k_0 | c_+|^2
 \ee
 as the incoming/outgoing  and absorbed/reflected fluxes.
  We define the amplification factor
  \be
Z = { \mathfrak{f}_{\rm out} \over  \mathfrak{f}_{\rm in} }  -1 =   {k_0\over p_3 |\tilde c_-|^2 } \left( |c_+|^2-  |c_-|^2 \right)
\ee
 that measures how much the incoming wave is amplified by the gravity solution. Super-radiance takes place when $Z>0$, i.e. when the ratio $k_0/p_3$ is positive for a black hole and negative for a perfectly reflecting ECO.

The absorption cross section is defined by the ratio
\be
\sigma_{abs} ={\mathfrak{f} _{\rm abs} \over \mathfrak{f} _{\rm in} }=  { k_0 | c_-|^2 \over p_3 |\tilde c_-|^2     }
\ee
The QNM's are defined as the poles of the absorption cross section $\sigma_{abs}$. They are  located at the zeros of $\tilde c_- =0$,.
The QNM frequencies are then defined as the solutions in $x_{\rm QNM}$  of the equation
  \be
  \mathfrak{R}={ c_+ \over c_- }= -{ M_{--} (x_{\rm QNM} ) \over M_{+-}(x_{\rm QNM} ) } \label{qnmeq}
  \ee
  for a given  $\mathfrak{R}$. For example the QNM's for a BH and a perfectly reflecting ECO are defined by
   \bea
 {\rm Black~ hole: } \quad & M_{--} = 0  \nn\\
  {\rm Perfectly ~reflecting~ ECO: } \quad & M_{+-} = 0
 \eea
  In the limit $b\to 0$, the BH condition $M_{--}=0$ can be written as
  \bea
  e^{ \partial_{p_2} \ln  \cF_{p_0}{}^{k_{0} }{}_{p_2}{}^{k}{}_{p_3}    (1 ,x)}  &=&  -{F^{-1}_{- +}[_{p_0 ~p_2}^{k_0~k_{12} }]
B_{+  -}[_{p_2^{+} ~p_3}^{k_{12}~k}]
 \over F^{-1}_{- -}[_{p_0 ~p_2}^{k_0~k_{12} }]
B_{-  -}[_{p_2^{-} ~p_3}^{k_{12}~k}]      } \\
&=& e^{2\pi {\rm i}  p_2}  {\Gamma( {-}2  p_2)^2  \over \Gamma(2  p_2)^2}
 \prod_{\alpha=\pm} { \Gamma(\ft12{+} k_0{+}\alpha p_0{+}p_2 ) \Gamma(\ft12{+} \alpha k_2{+} p_3{+}p_2 )
  \over \Gamma(\ft12{+} k_0{+}\alpha p_0{-}p_2 )\Gamma(\ft12{+} \alpha k_2{+}p_3{-}p_2 )      } \nn
  \eea
   There are infinite solutions to this equation parametrized by an integer, the overtone number. The results for the ECO can be obtained from those of the black holes after sending $k_0 \to - k_0$.

 \subsection{ Echoes   }

 ECO's are distinguished from black holes by their echo response to perturbations. We refer the readers to \cite{Mark:2017dnq} for a general discussion about
 echoes.  The radial wave response of a geometry to a localized perturbation is typically described (after Fourier transforming from time to the $\omega$ domain), by a non homogenous equation of the form
    \be
 \left[ {d^2\over dz^2} +Q(z) \right]  \Phi(z) = S(z) \label{inheq}
  \ee
 with $S(z)$  charactering the $t=0$ initial conditions (the shape of the perturbation)  localized someway inside the interval $z\in [0,1]$  and   vanishing  at the boundaries, i.e. $S(0)=S(1)=0$.
  We look for solutions to (\ref{inheq}) with boundary conditions
  \bea
  \Phi(z)  & \underset{z\to 1}{\sim}  f_-(z)   +
     \mathfrak{R} \,   f_+(z)  \qquad , \qquad
 \Phi(z)  & \underset{z\to 0}{\sim} \tilde f_+(z)    \label{greenbc}   \eea
 where $f_\pm(z)$, $\tilde f_\pm(z)$ are local solutions  and $\mathfrak{R}$  the  reflection coefficient.

 We denote by $\psi_{\rm in}(z)$, $\psi_{\rm out}(z)$ the solutions of the homogenous equation satisfying the boundary conditions
   \bea
   \psi_{\rm in} (z)  & \approx &
\left\{
\begin{array}{ccc}
   f_-(z)  &   & z\approx 1   \\
  A_+ \tilde f_+(z)  + A_- \tilde f_-(z)  &~~~~~~~~~~~~~~~~~~~~   & z\approx  0   \\
\end{array}
\right.
\nn\\
\psi_{\rm out} (z)  & \approx &
\left\{
\begin{array}{ccc}
 B_+ \, f_+(z)  + B_- \,  f_-(z)   & ~~~~~~~~~~~~~~~~~~~~  & z\approx 1   \\
   \tilde f_+(z)  & ~~~~  & z\approx  0   \\
\end{array}
\right.
   \eea
 with
 \be
A_\alpha=M_{-\alpha}
  \quad ,  \quad     B_\alpha=(M^{-1} )_{+\alpha}  \ee
 and Jacobian
 \be
  W=
{\rm det} \left(
\begin{array}{cc}
\psi_{\rm in} (z)   &  \psi_{\rm out} (z) \\
 \psi_{\rm in}' (z)   &  \psi_{\rm out}'(z)
\end{array}
\right)=
2  k_0 B_+ =2  p_3 A_- \quad ,\quad
 \ee
  Finally  the Green function $G(z,z')$, is defined by the equation
  \be
 \left[ {d^2\over dz^2} +Q(z) \right] G(z,z') = \delta(z-z')  \label{green}
 \ee
 with boundary conditions (\ref{greenbc}) in the $z$-variable.
   The Green function can be written as
 \be
  G(z,z') =
  \theta(z{-}z') {\psi_{\rm in}(z)  \psi_{\rm out}(z')  \over W }
  {+} \theta(z'{-}z) { \psi_{\rm in}(z')  \psi_{\rm out}(z) \over W } {+} { {\cal K} \over W } \psi_{\rm out}(z')  \psi_{\rm out}(z)
 \ee
 with
 \be
  {\cal K} ={   \mathfrak{R}   \over B_+ -  \mathfrak{R}  B_- }
    \ee
   The solution near $z\approx 0$ can therefore be written as
   \be
   \Phi(z) =\int dz' G(z,z') S(z') \underset{z\to 0}{\approx}
    { \tilde f_+(z)  \over W }  \left(  S_{\rm in}  +   {\cal K}   S_{\rm out}    \right)
   \ee
    with
    \be
    S_{\rm in} =\int \psi_{\rm in}(z')  S(z') dz'   \qquad , \qquad  S_{\rm out} =\int \psi_{\rm out}(z')  S(z') dz'
    \ee
     For a black hole, $\mathfrak{R}=0$ and therefore the echo ${\cal K}$-component is missing. For   $\mathfrak{R}$ small, a monochromatic  wave will
     therefore produce a sequence of echoes proportional to $\left( \mathfrak{R}  B_-/B_+\right)^n$ obtained by expanding the ${\cal K}$-signal for small $\mathfrak{R}$. On the other hand for a multi-chromatic signal, the Fourier transform of the echo signal will pick up either the poles of $W$ or the poles of ${\cal K}$. The former gives an echo signal with frequencies of the underlying black hole QNM's, while the latter gives
      an echo signal with the characteristic QNM's of the ECO given by the solutions to (\ref{qnmeq}).

 \section{Kerr-Newman geometries in four dimensions}

 In this section we compute the amplification factors and Love numbers for asymptotically flat Kerr-Newman BH's (or its associated ECO geometry) in four dimensions. The wave functions describing the radial and angular motion will be related to degenerate  five point conformal blocks in the confluent limit where
 two Fuchsian singularities collide.

%
%
\subsection{The wave equation}
The Kerr-Newman (KN) metric describes a solution to the Einstein-Maxwell equations in four dimensions with mass ${\cal M}$, charge ${\cal Q}$ and
angular momentum $J={\cal M} a_{\cal J} $.
 The line element of the KN metric in Boyer-Lindquist coordinates reads \cite{Caldarelli:1999xj}
\begin{equation}
\label{metric}
ds^2=-\frac{{\Delta_r}}{\rho ^2}(dt- a_{_\mathcal{J}}\sin^2\theta\, d\phi)^2 + \frac{\sin^2\theta}{\rho ^2}\left[a_{_\mathcal{J}}\,dt-
(r^2{+}a_{_\mathcal{J}}^2)\,  d\phi \right]^2 + \frac{\rho ^2 dr^2}{{\Delta_r}} + \rho ^2 d\theta^2
\end{equation}
where
\begin{equation}
\begin{aligned}
\Delta_r &=(r-r_+)(r-r_-)=r^2 -2 {\cal M} r +{\cal Q}^2+a_{_\mathcal{J}}^2\,,
\qquad
\rho^2 = r^2 + C^2\cos^2\theta  \,,
\end{aligned}
\end{equation}
These BH's posses two generally distinct horizons located at the zeroes of $\Delta_r$
\be
r_\pm  =  {\cal M}  {\pm} \sqrt{  {\cal M}^2{-}a_{_\mathcal{J}}^2{-}{\cal Q}^2 }
\ee
The wave-equation can be separated using the ansatz
\be
\Phi(t,r,\chi,\phi)=e^{{\rm i}(-\omega t+m_\phi \phi)}  R(r)S(\chi) =e^{{\rm i}(-\omega t+m \phi)}  \frac{ \widehat{R}(r)\widehat{S}(\chi)}{\sqrt{(1-\chi^2)\Delta_r(r)}}
\ee
with $\chi=\cos\theta$.
 In terms of these variables, the wave equation splits into two ordinary differential equations   of the form (\ref{can}) describing angular and radial motion.

\subsection{Angular motion}

In terms of the variables
\be
\Psi(z)=\widehat{S}(\chi),\quad \chi=\frac{2z}{ x}-1
\ee
The wave equation describing the angular motion takes the form  (\ref{can}) with
\bea
\label{qgravity}
\begin{aligned}
 Q_{\chi}(\chi) &=  \frac{(1-\chi ^2) (a_{_\mathcal{J}}^2 \omega ^2\chi ^2+A)-m^2+1}{(1-\chi^2)^2}\label{qswkn}
 \end{aligned}
 \eea
where $A$ is the separation constant and $m$ the azimuthal angular momentum of the incoming wave.
 Matching (\ref{qgravity}) against the BPZ $Q$-function
\bea
Q^{\rm angular}_{1,2}(z) = \frac{\delta _{k_2}}{(x-z)^2}+\frac{\delta_{p_3}}{z^2} -
\frac{c^\chi}{z} +\frac{u-\frac{1}{4}+\delta _k+\delta _{p_3}
+ x\left(k_2+\ft12 \right)}{z (x-z)} -\frac{1}{4}
\eea
one finds the gauge/gravity dictionary
\bea
x=q^{\chi}= 4 a_{_\mathcal{J}} \omega\,,\,\, c^{\chi}=0\,,\,\, k_2^{\chi}=p^{\chi}_3=\frac{m}{2}\,,\,\,
u^{\chi}=A{+}\frac{1}{4}{+}a_{_\mathcal{J}}^2 \omega ^2{-}2 a_{_\mathcal{J}}\omega  (m+1)\label{gaugegravityang0}
\eea
where the superscript $\chi$ is to distinguish the angular variables from the radial ones in the next section.
In the gauge theory language regularity of the wavefunction $\hat S(\chi)$ at the boundaries $\chi=\pm 1$ requires the  $a$-cycle quantization condition \cite{Bianchi:2021xpr,Bianchi:2021mft}
\be
p^{\chi}_2=a =  \ell+\ft12 \label{pchi}
\ee
with $\ell$ a non-negative integer. When $ a_{_\mathcal{J}}  \omega=0$, the gauge theory is free ($q=0$) and the equation can be explicitly solved in terms of hypergeometric functions (spherical harmonics)
\be
\Psi(\chi)\underset{a_{_\mathcal{J}}  \omega=0}{\approx} \sqrt{1-\chi^2} P_{\ell m}(\chi) ={\sqrt{1-\chi^2}\over \Gamma(1-m)}\left( {1+\chi\over 1-\chi}\right)^{m\over 2} {}_2 F_1\left(-\ell,\ell+1,1-m;{1-\chi\over 2}\right)
\ee
with $m$ an integer with $\ell>|m|$. The separation constant becomes
\be
A \underset{a_{_\mathcal{J}}  \omega=0}{\approx}  \ell(\ell+1)
\ee
 In this limit, (\ref{uqsmall}) and (\ref{gaugegravityang0}) yield  $u^{\chi}\approx  (a^{\chi})^2\approx (\ell+\ft12)^2$. Turning on the contribution of $a_{_\mathcal{J}}  \omega$, the relation between $u$ and $a$ is determined order by order in $q^\chi$ from the difference equation  (\ref{fractionequality})
 with
\bea
  P_0(a) &=& -(a-a_{02}) \quad, \quad  P_3(v) =(a-a_{31})(a-a_{32}) \nn\\
P_2(a) &=& a^2-u^\chi +q^\chi  (a_{02}+a_{31}+a_{32} - \hbar -a)
\eea
Plugging the ansatz (\ref{uqsmall}) for $u^\chi(a)$ into (\ref{fractionequality}), solving for the expansion coefficients and using the gauge gravity dictionary (\ref{gaugegravityang0}) one finds
{\small
\bea
&&A=  \ell(\ell+1) +{a_{\cal J}^2 \omega^2 (1+2 m^2-2 \ell(\ell+1) ) \over 4\ell(\ell+1)-3} \label{aomega} \\
&&+\frac{2 a^4 \omega ^4 \left[4 \ell^6{+}12 \ell^5{-}3 \ell^4{-}26 \ell^3{+}2
   \ell^2{+}17 \ell{-}3{+}\left(20 \ell^2{+}20 \ell{+}33\right) m^4-\left(24 \ell^4{+}48 \ell^3{-}2 \ell^2{-}26 \ell{+}30\right) m^2\right]}{(2 \ell{-}3) (2 \ell{+}5) (2 \ell{+}3)^3(2 \ell{-}1)^3}+\ldots \nn
\eea
}
 with the leading correction reproducing the result (12) in \cite{Starobinskil:1974nkd} in the scalar case after the identification
 ${}_0 \lambda_l^{m_\phi}=A+a_{\cal J}^2 \omega^2$ and (1.4) in \cite{Mano:1996vt}.  The $\omega^4$ term has not appeared previously in the literature.

\subsection{Radial motion}

The radial equation can be written in the (\ref{can}) form with $\widehat{R}(r ) =z^{-1}\Psi(z)$ and
\bea
\label{qgravity2}
\begin{aligned}
 Q_{r} (r) &= \frac{\left( \omega (r^2+a_{_\mathcal{J}}^2) - a_{_\mathcal{J}} m \right)^2-\Delta _r \left( a_{_\mathcal{J}}^2 \omega^2 -2 a_{_\mathcal{J}} \omega m+A+1 \right)+\frac{1}{4} \Delta _r^{'}{}^2   }{\Delta _r^2}
 \end{aligned}
 \eea
 This function displays two Fuchsian singularities at the horizons $r=r_\pm$ and an irregular singularity at $r=\infty$. Using the coordinates
  \be
 z={ r_+- r_- \over r-r_-}   \label{zrr}
 \ee
we map the region $r\in [r_+,\infty]$ into the interval $z\in [0,1]$ with $1,0$ corresponding to the outer horizon and infinity respectively.
Plugging  (\ref{zrr}) into (\ref{qqschwarzian}) one finds that $Q(z)$ has the form of (\ref{Qijns})
 \be
\label{Q21ns}
Q^{\rm radial}_{2,1}(z)= \frac{ \delta_{p_0}}{z (z{-}1)}{+}
\frac{\delta_{k_0}}{(z{-}1)^2 z} {+}  \frac{u-\frac{1}{4} +x\left(\frac{1}{2}-k_0 \right)}{z^2(z-1) } {+}
\frac{c\, x }{z^3}-{x^2\over  4z^4}
\ee
 with gauge and gravity variables identified according to
 \bea
&&x =q= 2{\rm i}  \omega (r_+-r_-)~, ~\,
k_0= {\rm i} Q_{\rm grav}   ~, ~\, p_0 ={\rm i} Q_{\rm grav} +2 {\rm i} \omega {\cal M}  ~, ~\,  c ={-}2{\rm i}\omega {\cal M} \nn \\
&&u  =A{+}\ft{1}{4}{-}{\rm i}  \omega (r_+{-}r_-)(1{-}2{\rm i} Q_{\rm grav})  {-}\omega^2(a_{_\mathcal{J}}^2{+}4{\cal M}^2 {+}2 r_+^2)
 \label{gaugegravityrad2}
\eea
with
\be
 Q_{\rm grav}= { a_{_\mathcal{J}} m - \omega (r_+^2+a_{\cal J}^2)   \over r_+ -r_-} =  ( \Omega  m - \omega)  { r_+^2+a_{\cal J}^2 \over r_+ -r_-}
 \label{qgrav}
 \ee
where $\Omega$ is coordinate angular velocity. For $\omega \approx 0$, $\sqrt{u} \approx \ell+\ft12$, and the ingoing solution $\Psi_-$ can be written in the analytic form
\be
\Psi_-(z)\underset{ \omega= 0}{\approx}   c_\alpha \left( 1{-}z\right)^{ {1\over 2}{ }-k_0} z^{ {1\over 2} {+}\alpha \sqrt{u}  }  {}_2
F_1\left(\ft12{+}\alpha \sqrt{u}{-}k_0{-}p_0, \ft12{+}\alpha \sqrt{u}{-}k_0{+}p_0, \ft12{+}2\alpha \sqrt{u},  z\right)
\ee
 with $c_\alpha$ some constants chosen such that the  term falling as $(1-z)^{{1\over 2} -k_0}$  vanishes. 
 For $\omega \neq 0$, the solution can be written as an infinite sum of hypergeometric functions (whose indices and coefficients  computable order by order in $\omega$) \cite{Mano:1996vt}. We can check these results against the gauge theory formula for $\Psi_\alpha(z)$ given in (\ref{connection}).
 We are interested in solutions with ingoing boundary conditions at $z\approx 1$, so we focus on the component
   $\Psi(z) \sim  \Psi_-(z)$. Again the non trivial task is to determine the relation between $u$ and $ p_2$.
 As before, this can be done using (\ref{fractionequality}) where now
\bea
  P_0 (a)&=& (a-a_{01})(a-a_{02}) \quad, \quad  P_3(a) =-(a-a_{32}) \nn\\
P_2(a) &=& a^2-u +q (a_{01}+a_{02}+a_{32}  -a)
\eea

    Solving (\ref{fractionequality}) now for $a(u)$  order by order in $q$,
    using the gauge gravity dictionary (\ref{gaugegravityrad2}) and
   formula (\ref{aomega}) for the separation constant one finds
   \bea
    p_2 &=& a=\ell+\ft12+ \Delta \nu =\ell+\ft12 +    2 \omega^2 {  {\cal M}^2  \left[ 11{-}15\ell(\ell{+}1) \right]  {+} {\cal Q}^2 \left[ 3\ell(\ell{+}1) -2\right]      \over (2\ell-1)(2\ell+1) ( 2\ell+3) }  \nn\\
    &&+\frac{4\, a\, m\, \omega^3 \left(2 \left(5 \ell^2+5 \ell-3\right) {\cal M}^2-\ell (\ell+1)
   {\cal Q}^2\right)}{\ell (\ell+1) (2 \ell-1)(2 \ell+1) \left(2\ell+3\right)} +\ldots \label{p2corr}
   \eea
with the leading correction matching (5.2) in  \cite{Mano:1996vt} for $Q=s=0$.  The $\omega^3$ term has not appeared previously in the literature.

\subsection{Love numbers }

Love and dissipation numbers parametrize the deformability of a geometry under an external perturbation.
 They can be extracted from the asymptotic behaviour at infinity of a scalar wave propagating in the gravity background. For a static wave, $\omega=0$,  the Love number is defined as the ratio between the term falling as $r^{-\ell-1}$ and the one growing as $r^\ell$ representing the response and the source terms respectively. These two terms can be unambiguously distinguished if we take $\ell$ to be an almost (but not exactly) an integer.
  When $\omega \neq 0$, the exponents of the two solutions get $\omega$ corrections and there is no (as far as we know) obvious definition of the Love number.  Here, we propose a gauge theory inspired definition of the Love number, given again by the ratio between the leading coefficients
  of the solution now computed in the double scaling limit  where both $z$ and $x/z$ (the two gauge couplings) are small, i.e.
  \be
   z\approx  \ft{r_+-r_-}{r} <<1  \qquad , \qquad \frac{x}{z}\approx \omega r <<1
  \ee
or equivalently
  \be
r_+- r_- << r << \omega^{-1}
\ee
 In this limit the wave function can be written as
    \be
 \Psi(z) \underset{z\to 0 }{\sim}     z^{-\ell -\Delta \nu} +  z^{\ell+1+\Delta\nu }   \mathfrak{L}_0 \,  \left[ 1+ \ldots \right] \label{phil00}
 \ee
  and the dynamical Love  $L$ and dissipation number $\Theta$ can be identified as the real and imaginary parts of
  the ratio $\mathfrak{L}_0$, i.e.
  \be
  \mathfrak{L}_0 =L+{\rm i} \Theta
  \ee
  In the case under study, the solution with  ingoing boundary conditions at $z\approx 1$ is proportional to $\Psi_-(z,x)$ in (\ref{connection}).
   From (\ref{psiminus00}) one finds
   \be
\Psi_- (z )   \underset{ z\to 0}{\approx}  e^{{-}{x\over z } }  \sum_{\alpha'}  z^{{1\over 2} + \alpha'  p_2 }    F^{-1}_{-  \alpha'}[_{p_0 ~p_2}^{ k_0 ~k_{12} }]
   {}_1 F_1 (  \ft{1}{2}{+} c {-}\alpha' p_2,
    1{-}2 \alpha'  p_2, \ft{2 x}{z} )  \label{psiminus0}
\ee
  with
 \be
 F^{-1}_{- \alpha'}[_{p_0 ~p_2}^{k_0~k_{12} }]   = \frac{\Gamma (1+2   k_0)
\Gamma \left( -2  \alpha' p_2  \right)}{\Gamma \left(  \ft12 -\alpha' p_2  + k_0+p_0
\right) \Gamma \left( \ft12-\alpha' p_2 + k_0-p_0  \right)} \label{fexp}
\ee
  Expanding for $z\to 0$ and computing the ratio between the two terms $\alpha'=\pm$, one finds the remarkable simple formula
   \be
\mathfrak{L}_0 =L+{\rm i} \Theta  ={  F^{-1}_{- +}[_{p_0 ~p_2}^{ k_0 ~k_{12} }]
     \over
  F^{-1}_{- -}[_{p_0 ~p_2}^{ k_0 ~k_{12} }]
   }=    \frac{\Gamma \left( -2  p_2  \right)
  \Gamma\left(\ft12 {+}  k_0{+}p_0+p_2 \right) \Gamma \left(
   \ft12 {+} k_0 {-} p_0+p_2 \right) }{   \Gamma \left( 2   p_2 \right) \Gamma
   \left(\ft12 + k_0{+} p_0-p_2 \right) \Gamma \left(
  \ft12 {+}  k_0 {-} p_0-p_2 \right)} \label{love1}
\ee
with $p_0,k_0$ given by (\ref{gaugegravityrad2}) and $p_2$ by (\ref{p2corr}).
(\ref{love1}) reproduces the standard results for the Love and dissipation number of Kerr-Newman BH’s. Indeed, In this limit  $p_0 = k_0 = {\rm i} Q_{\rm grav}$, $p_2=\ell+\ft12$, with $\ell$ an almost (but not exactly) an integer, leading to
\be
\mathfrak{L}_0 \approx    \frac{\Gamma \left( {-}2\ell {-}1\right)\Gamma \left( \ell {+}1 \right)
  \Gamma\left(\ell {+}1  {+}2 {\rm i} Q_{\rm grav}   \right)  }{   \Gamma \left( 2 \ell {+}1 \right)\Gamma \left(
   - \ell  \right) \Gamma
   \left( -\ell  +2 {\rm i} Q_{\rm grav} \right) }=
  -  \frac{  {\rm i} Q_{\rm grav}\,  \left( \ell\right)!^2 }{   \left( 2 \ell \right)!  \left( 2 \ell +1\right)!   } \prod_{n=1}^\ell \left( n^2 +  4 Q_{\rm grav}^2   \right)\label{love}
\ee
in agreement with (3.55) of  \cite{Charalambous:2021mea} (or (113) of \cite{Bonelli:2021uvf}.

 Finally, we observe that the limit $\omega r<<1$ can be relaxed and  the exact dependence on $\omega r$ of the ratio  between the response
 and the source terms can be explicitly determined from (\ref{psiminus0}). Indeed,  the wave function at infinity can be written as
  \be
 \Psi(z) \underset{z\to 0 }{\sim}     z^{-\ell-\Delta \nu} +  z^{\ell+1+ \Delta\nu}   \mathfrak{L}( \ft{x}{z})\,  \left[ 1+ O(z) \right]
 \ee
  with $\mathfrak{L}( \ft{x}{z})$ a sort of ``Love function" given by
 \be
   \mathfrak{L}(\ft{x}{ z})  =\mathfrak{L}_0 {
 {}_1 F_1 (  \ft12+c -p_2 ,  1-2 p_2 , \ft{2x}{z} )   \over
    {}_1 F_1 (  \ft12+c+p_2 ,  1+2 p_2 , \ft{2x}{z} ) }
 \ee
   It would be interesting to see whether the first expansion coefficients of this function in powers of $\omega r$ can be detected in the GW signal.

\subsection{Amplification factors}

In  this section we compute the amplification factors for the KN BH in the slowly rotating limit $\omega {\cal M} <<1$.
The asymptotic solution at infinity is given by  (\ref{allsolutions2}) \be
   \Psi_-(z)      \underset{z\to 0}{\sim}     Y_{-}  z^{1-c} e^{- { x\over z}}  +
 Y_{+} \, z^{1+c}\, e^{ { x\over z}}
 \label{psiminus}
   \ee
  with
    \be
 Y_{\alpha''} =  \sum_{\alpha'} {\cal Y}_{\alpha' \alpha''} = \sum_{\alpha'}
 x^{\alpha' p_2 -  \alpha'' c  } F^{-1}_{- \alpha'}[_{p_0 ~p_2}^{k_0~k_{12} }]  \,B^{\rm conf}_{\alpha' \alpha''}[_{p_2 ~c}^{k_{12}}]
 \label{yalpha0}
 \ee
 and $F^{-1}$,  $B^{\rm conf}$ given by (\ref{fexp}) and (\ref{braidingBconf}).
 Explicitly
  \be
{\cal Y}_{\alpha' \alpha''} =   \frac{x^{\alpha' p_2 -  \alpha'' c  }  \, e^{-{\rm i} \pi \delta_{1\alpha''}({1\over 2} {+} c{-} \alpha' p_2)} \, \Gamma (1+2   k_0)
\Gamma \left( -2  \alpha' p_2  \right) \Gamma( 1- 2\alpha' p_2 )}{\Gamma \left(  \ft12   + k_0+p_0-\alpha' p_2
\right) \Gamma \left( \ft12 + k_0-p_0 -\alpha' p_2 \right)\Gamma( \ft12 -\alpha'' c-\alpha' p_2) } \label{fexp2}
\ee
 In terms of these variables the amplification factor becomes
\be
 Z  = 1- \left| { Y_{+} \over Y_{-} }\right|^2 =  1- \left| { {\cal Y}_{-+}+{\cal Y}_{++} \over  {\cal Y}_{--}+{\cal Y}_{+-}   }\right|^2 \label{zamplif0}
 \ee
 In the slowly rotating limit $x \to 0 $, $p_0 \approx k_0  \approx  {\rm i } Q_{\rm grav}$, $c\approx 0$,  $p_2\approx \ell+\ft12$  with $\ell$ almost (but not exactly) an integer. One finds
 \be
 Z  \approx 1- \left| {  1-\Sigma \over  1+\Sigma }\right|^2   \approx 4 {\rm Re} \Sigma \label{amplif}
 \ee
with
  \be
\Sigma={ {\cal Y}_{+-} \over {\cal Y}_{--} }  \approx   - \frac{x^{2\ell+1   }   \,
  \Gamma( - 2 \ell )^2 \Gamma \left( \ell+1 \right)^2  \Gamma \left(  \ell+1+2 {\rm i } Q_{\rm grav}
\right)}{
\Gamma \left(   -\ell \right)^2  \Gamma( 2 \ell+2 )^2 \Gamma \left(     -\ell+2 {\rm i } Q_{\rm grav}
\right)  } \label{fexp2}
\ee
  leading to
 \be
 Z     \underset{x\to 0}{\approx}         { 2\, |x|^{2\ell+1}  \,Q_{\rm grav} \ell !^4 \over  (2\ell)!^2 (2\ell+1)!^2}     \prod_{n=1}^\ell
   \left( n^2{+} 4 Q_{\rm grav}^2  \right) \label{amplif4}
   \ee
  in agreement with formula (26) in \cite{Starobinskil:1974nkd}. Superradiance takes place when $\omega <m_\phi \Omega   $.
  For the perfectly reflecting ECO, one finds the same result with $k_0 \to -k_0$, i.e. $Z \to -Z$.
   Superradiance in this case is obtained when $Q_{\rm grav}<0$ i,e. when $\omega >m_\phi \Omega   $ !

\section{Solutions of the Einstein-Maxwell theory in five dimensions}

The results in the last section can be easily generalized to BH in five dimensions, since as shown in \cite{Bianchi:2021xpr,Bianchi:2021mft,Bianchi:2022wku,Bianchi:2021yqs},
they can be described by different confluent limits of the same ordinary differential equation.

\subsection{The wave equation}

 We consider the Einstein-Maxwell theory with Lagrangian
\be
16\pi G_5 \,{\cal L}_{5d} =\sqrt{g_{5d}} \left (R_{5d}
 -{1\over 4} F^2 \right)-{1\over 3\sqrt{3}} F\wedge F \wedge A
\ee
 We set  units such that the five dimensional Newton constant\footnote{The constant $G_5$ has been chosen so that the expansion at large distance of the $g_{tt}$ component of the metric reads  $g_{tt} \sim -1 + \frac{2\cal M}{r^2}$.} is $G_{5} = 3\pi/4$.  The general solution obtained in \cite{Chong:2005hr}
 is specified by four parameters, the mass ${\cal M}$, the charge ${\cal Q}$ and two angular momentum parameters  ${\ell_1}$, ${\ell_2}$. The line element reads
\begin{equation}
\begin{aligned}
ds^2 &= -dt^2-\frac{2 \mathcal{Q} \,\omega _2 }{\Sigma }(dt-\omega _1)+\Delta _t(dt-\omega_1)^2  +
\Sigma  \left(d\theta^2+\frac{r^2dr^2}{\Delta _r}\right) +
\\
&
 + d\psi^2 \cos^2\theta \,(r^2+\ell _2^2 )+d\phi^2 \sin ^2\theta \,(r^2+\ell _1^2)
\end{aligned}
\end{equation}
with
\bea
\begin{aligned}
\omega _1 &= \ell_2 \,\cos ^2 \theta\,d\psi + \ell_1 \,\sin ^2\theta\, d\phi
\quad , \quad
\omega _2  = \ell_1 \,\cos ^2 \theta\,d\psi + \ell_2 \,\sin ^2\theta\, d\phi  \nn\\
\Delta _r &= (r^2 - r_+^2)(r^2 - r_-^2)
\, , \quad
\Delta _t = \frac{2 \mathcal{M} \Sigma -\mathcal{Q}^2}{\Sigma ^2}
\, , \quad
\Sigma = r^2+\ell _1^2\cos ^2\theta+\ell _2^2\sin ^2\theta
\end{aligned}
\eea
These geometries possess two horizons located at
\begin{equation}
y_\pm =r_\pm^2 =\widehat{{\cal M}} \pm \sqrt{\widehat{{\cal M}}^2 - \widehat{{\cal Q}}^2}
\end{equation}
with
\begin{equation}
\widehat{{\cal M}} = {\cal M} - \frac{\ell_1^2+\ell_2^2}{2}
\quad , \quad
\widehat{{\cal Q}} = {\cal Q} + \ell_1 \ell_2
\end{equation}
Introducing the variables $\chi=\cos\theta$, $y=r^2$, the scalar wave equation  can be separated, setting
\be
\Phi(t,y,\chi,\phi_1,\phi_2)=e^{{\rm i}(-\omega t+m_\phi \phi +m_\psi \psi)}  R(r)S(\cos\theta) =e^{{\rm i}(-\omega t+m_\phi \phi +m_\psi \psi)}  \frac{\widehat{R}(y) \widehat{S}(\chi)}{\sqrt{\Delta_\chi(\chi) \Delta_y(y)} }
\ee
with
\be
\Delta_y=(y-y_+)(y-y_-). \qquad , \qquad \Delta_\chi(\chi) =\chi(1-\chi^2)
\ee

\subsection{Angular motion}

The angular equation can be written in the normal form (\ref{can}) for $\widehat{S}(\chi)$ with
 \begin{equation}
\begin{aligned}
Q_{\chi } (\chi) &= \frac{1{-}4(1{-}\chi^2) m_{\psi }^2{-}4\chi^2  m_{\phi }^2  {+}4  \chi^2(1{-}\chi^2)  \left(A{+}\omega ^2 \chi^2 \left[ \ell _1^2{+} \ell _2^2 (1{-} \chi^2)\right]+3\chi^2(1{-}\chi^2) \right)}{4  \chi^2(1-\chi^2) }\label{angular5}
\end{aligned}
\end{equation}
and $A$ the separation constant. The $Q_{\chi } $-function displays Fuchsian singularities at $\xi=0,1$ and an irregular singularity at infinity, so
writing $\widehat{S}(\chi)=(4x-4z)^{-{1\over 4}} \Psi(z)$ and
\be
\chi^2=1-{z\over x}
\ee
the differential equation takes the normal form with
\be
Q_{0,2} = \frac{\delta _{k} }{(x-z)^2}+\frac{\delta_{p_3}}{z^2}-
\frac{1}{z} +\frac{ u-\frac{1}{4} +\delta _{k} +\delta _{p_3}}{z (x-z)}
\ee
and
\be
x=q^\chi={\omega^2\over 4} (\ell_2^2-\ell_1^2) \quad , \quad p^\chi_3 ={m_\phi\over 2} \quad , \quad k^\chi={m_\psi\over 2} \quad, \quad
u^\chi=\ft14( A+1+  \omega^2\,\ell_2^2 ) \label{diccclp0}
\ee
  We notice that for $\ell_1^2=\ell_2^2$, the gauge theory is free and therefore the differential equation can be solved in terms of hypergeometric functions.
  For simplicity, we will restrict to this case, so we will write
  \be
   p_2^\chi  =\sqrt{u^\chi} =\sqrt{A+1\over 4} ={\ell+1 \over 2}
  \ee
  i.e.    $A= \ell(\ell+2)$.
   In terms of these variables, the two solutions read
  \be
  \Psi_\alpha  =   (z{-}x)^{m_\psi{+}1\over 2} \left( {{\rm i} z\over x} \right)^{1+\alpha m_\phi \over 2} {}_2 F_1 ( 1{+}\ft12( m_\psi {+}\alpha m_\phi {+}\ell),
  \ft12( m_\psi {+}\alpha m_\phi {-}\ell), 1{+}\alpha m_\phi, \ft{z}{x})
  \ee
  with $\alpha=\pm$.
  Finally requiring regularity at $z\approx x$, i.e. the cancelation of the  $\log(z-x)$-term, one finds a single regular solution (hyperspherical harmonics on $S^3$) assuming $\ell$,  $m_\phi$,
  $m_\psi$ are all quantized.

\subsection{Radial motion}

Now let us consider the radial motion.  It is convenient to introduce the following combinations
\bea
m_+ &=& (m_\phi+m_\psi) (\ell_1+\ell_2)-\omega  (2
   \mathcal{M} - {\cal Q} ) \nn\\
  m_- &=& (m_\phi-m_\psi) (\ell_1-\ell_2)-\omega  (2
   \mathcal{M}+{\cal Q}) \nn\\
 2\kappa &=& \widehat{{\cal M}}  (m_+-m_-)^2+ \widehat{{\cal Q}}  (m_+^2-m_-^2)
   \eea
    parametrizing the frequencies $\omega$, and the two angular quantum numbers $m_\phi$, $m_\psi$.
 In terms of these variables, the radial wave equation can be written in the form (\ref{can}) for $\widehat{R}(y)$  with
\begin{equation}
\begin{aligned}
Q_{y} (y) &= \frac{y\,  m_+ m_- +\kappa -\Delta _y \left(A+4-\omega ^2 \left(y+2 \mathcal{M}   \right)\right)+\Delta _y'{}^2}{4 \Delta _y^2}
\end{aligned}
\end{equation}
 and $A$ the separation constant. Introducing the variable
\be
z={y_+ - y_-\over y-y_- }
\ee
one can write the radial wave equation in its normal form (\ref{can})
with
 \be
Q_{2,0}(z)=  \frac{ \delta_{p_0}   }{z (z{-}1)}{+}   \frac{\delta_{k_0}  }{(z{-}1)^2 z} {+}  \frac{ u-\ft{1}{4}  }{z^2(z-1) } {-}
 \frac{ x }{z^3}
\ee
and
\bea
x &=& q  = {\omega^2 \over 4} (r_+^2 {-} r_-^2) \qquad ,\qquad u=\ft14\left( A+1- 3  {\cal M} \omega^2 +\omega^2   (r_-^2+2 {\cal M} )  \right) \nn\\
k_{0}  &=&{\rm i} Q_{\rm grav} ={\rm i} {  \sqrt{r_-^2 m_+ m_-  {+}\kappa} \over 2 (r_+^2{-}r_-^2)}   \qquad ,\qquad p_{0}  ={\rm i}  {  \sqrt{r_+^2 m_+ m_-  {+}\kappa} \over 2 (r_+^2{-}r_-^2)} ={\rm i} Q_{\rm grav} + \Delta p  \label{diccclp}
\eea
In the limit $\omega {\cal M}$ small, the relation between $p_2$ and $u$ is determined from  (\ref{fractionequality})
 with
\bea
 P_0(v) =(v-a_{01})(v-a_{02} )  \qquad , \qquad P_3(v)=1  \quad, \quad  P_2(v) = v^2-u +q
\eea
 Solving for $a(u) $ order by order in $q$, and using the dictionary   (\ref{diccclp}) one finds
 \be
 p_2 ={\ell+1\over 2}  + \Delta \nu ={\ell+1\over 2}  -{\omega^2 ( m_+ m_- +\ell(\ell+2)(r_+^2+r_-^2+4 {\cal M})   ) \over 8\ell(\ell+1)(\ell+2) } +\ldots   \label{p2corr2}
 \ee

\subsection{Love numbers }

 The computation of Love and dissipation numbers in five dimensions follows {\it mutatis mutandis} the same steps as before.
 The asymptotic solution is given again by (\ref{psiminus0}) leading to
  \be
 \Psi(z) \underset{z\to 0 }{\sim}     z^{-{\ell\over 2} -\Delta \nu} +  z^{ {\ell\over 2} +1+\Delta\nu }   \mathfrak{L}_0 \,  \left[ 1+ \ldots \right] \label{phil5}
 \ee
  The Love number is given again by the universal formula
 (\ref{love1}) where now  $p_0$, $k_0$ and $p_2$  are given by (\ref{diccclp}) and (\ref{p2corr2}).
 In the slowly rotating limit $p_2 \approx \ft{\ell+1}{2}$, $p_0=k_0={\rm i} Q_{\rm grav}$  leading to
 \be
\mathfrak{L}_0 \approx  -  \frac{\Gamma \left( -\ell  -2\Delta \nu \right)\Gamma \left( \ft{\ell}{2}+1 \right)
  \Gamma\left(\ft{\ell}{2} +1  +2 {\rm i} Q_{\rm grav}   \right)  }{   \Gamma \left(  \ell+2 \right)\Gamma \left(
   - \ft{\ell}{2}  \right) \Gamma
   \left( -\ft{\ell}{2}  +2 {\rm i} Q_{\rm grav} \right) } \label{love5}
\ee
  For $\ell$ even one finds
   \be
\mathfrak{L}_0 =L+{\rm i} \Theta   \approx
    -
    \frac{  {\rm i} Q_{\rm grav}\,  \left( \ft{\ell}{2}\right)!^2 }{   \left(  \ell \right)!  \left(  \ell +1\right)!   } \prod_{n=1}^{\ft{\ell}{2}} \left( n^2 +  4 Q_{\rm grav}^2   \right)\qquad \ell\in 2 \mathbb{N}
    \label{love51}
\ee
For $\ell$ odd, $\mathfrak{L}_0$ diverges, and the static limit is not well defined due to the mixing of the source and the response terms.
The leading $\log z$ term in the expansion of (\ref{phil5}) is instead finite
\be
 \Psi(z) \underset{z\to 0 }{\sim}     z^{-\ell} +  z^{\ell+1} (  \mathfrak{L}_0  + R \log z+\ldots ) \,  \left[ 1+ O(z) \right]
 \ee
with
\be
R = 2 \,\Delta \nu \, \mathfrak{L}_0 =   \frac{ (-)^\ell \Gamma \left( \ft{\ell}{2}+1 \right)
  \Gamma\left(\ft{\ell}{2} +1  +2 {\rm i} Q_{\rm grav}   \right)  }{ \ell!   \Gamma \left(  \ell+2 \right)\Gamma \left(
   - \ft{\ell}{2}  \right) \Gamma
   \left( -\ft{\ell}{2}  +2 {\rm i} Q_{\rm grav} \right) } \qquad \ell\in 2 \mathbb{N}+1   \label{love52}
\ee
  (\ref{love51}) and  (\ref{love52}) reproduces (60) of \cite{Pereniguez:2021xcj} after setting $Q_{\rm grav} \to 0$ and $\ell \to 2 l$.

\subsection{Amplification factors}

The computation of the amplification factor follows the same steps that the one in four dimensions. The asymptotic solution is given again by
(\ref{psiminus}), and the amplification factor by (\ref{zamplif0}) and (\ref{yalpha0}) with
  $p_0$, $k_0$ and $p_2$  given by (\ref{diccclp}) and (\ref{p2corr2}). Now it makes difference if $\ell$ is even or odd.
  For example, in the static limit $\omega \to 0$, for $\ell$ even the amplification factor is given again by  (\ref{amplif4}) with $\ell \to \ell/2$
   \be
 Z     \underset{x\to 0}{\approx}         { 2\, |x|^{\ell+1}  \,Q_{\rm grav} (\ft{\ell}{2}) !^4 \over  (\ell)!^2 (\ell+1)!^2}     \prod_{n=1}^{\ell\over 2}
   \left( n^2{+} 4 Q_{\rm grav}^2  \right) \qquad \ell\in 2 \mathbb{N}  \label{amplif5}
   \ee
while for $\ell$ odd one finds  $Z     \underset{x\to 0}{\approx}    1$.

\section{Summary of results}

 In this work, we have studied the scattering of waves from BH's and ECO's geometries using conformal field theory and localization techniques. Following
 \cite{Bonelli:2021uvf}, we related the
   wave functions to the conformal blocks of Liouville theory and the connection matrices  to "braiding" and "fusion" moves of the CFT.

  We  applied the results to the study of waves propagating in the background of BH's and ECO's in four and five dimensions. ECO's are  represented by BH geometries ending on a membrane with reflectivity properties parametrized by the coefficient $\mathfrak{R}$. The QNM's are defined as poles (in the space of frequencies) of the absorption cross section and they are given by the solutions of the equation
  \be
  M_{--}+\mathfrak{R} M_{+-}=0
  \ee
  with $M_{\alpha \alpha''}$ the connection matrix relating the solution at the would be horizon and infinity.  The same characteristic frequencies appear as poles in the echo response  of the geometry.

  We studied also the tidal response of BH's and ECO's  to perturbations slowly varying with time. In four dimensions, the tidal response was extracted from the asymptotic form
  of the scalar wave function in the region $r_+-r_- << r << \omega^{-1}$.  We write
  \be
 \Psi(r) \underset{r\to \infty }{\sim}    1 +   \left({r\over r_+ {-} r_-}\right)^{-2\ell-1-2 \Delta\nu  }  \left[  \mathfrak{L}(\omega r)
+   \ldots \right]   \label{phil0}
 \ee
  with $\Delta\nu$ given by (\ref{p2corr}) and dots denoting $\ft{r_+-r_-}{r}$ corrections. The ratio between the response and source terms is parametrized by a  ``Love function"  $\mathfrak{L}(\omega r) $ (rather than a number). We obtained an exact formula for this function in terms of hypergeometric functions.
   A similar formula {\it mutatis mutandis} is found in five dimensions with $r \to r^2$, $\ell \to \ell/2$.
  We define  the dynamical Love $L$ and dissipation $\Theta$  numbers as the real and imaginary parts of $\mathfrak{L}_0=\mathfrak{L}(0) $.
   We obtained the universal formula
    \be
 \mathfrak{L}_0 =L+{\rm i} \Theta =
  -    \frac{\Gamma \left( -2  p_2  \right)
  \Gamma\left(\ft12 {+}  k_0{+}p_0+p_2 \right) \Gamma \left(
   \ft12 {+} k_0 {-} p_0+p_2 \right) }{   \Gamma \left( 2   p_2 \right) \Gamma
   \left(\ft12 + k_0{+} p_0-p_2 \right) \Gamma \left(
  \ft12 {+}  k_0 {-} p_0-p_2 \right)}  \label{generalloveF}
     \ee
   valid for all BH's in four and five dimensions.  The parameters $k_0$, $p_0$ and $p_2$  are given by  (\ref{gaugegravityrad2}) and (\ref{p2corr}) for Kerr-Newmann in four dimensions, and by (\ref{diccclp}) and (\ref{p2corr2}) in five dimensions. The non-triviality of the differential equation is codified into a single function: the quantum SW period $p_2=a(u)$ specifying the gauge theory prepotential ! This function can be derived from a continuous fraction version of the quantum deformed SW curve  \cite{Poghossian:2010pn,Poghosyan:2020zzg,Fucito:2011pn}. In the $\omega \to 0$ limit, (\ref{generalloveF}) reproduces the known  results for the static Love number of BH's in four and five dimensions.

   We conclude with a comment on a subtle point in the definition of the static Love number. We recall that the static Love number is well defined only if we analytically continue $\ell$ to be almost (but not exactly) an integer. This analytic continuation of $\ell$ is required in order to have an unambiguous split between the source $r^\ell$ and  the response $r^{-\ell-1}$ terms when $\omega $ is exactly zero.  When $\omega \neq 0$ there is no such mixing, so one could do without this analytic continuation in $\ell$. Remarkably the two limits  $\omega\to 0$ and $\ell \to \mathbb{N}$ do not commute !
  Indeed sending $\ell \to \mathbb{N}$ and then  $\omega\to 0$  yields a non-trivial result for the Love number that grows with $\omega^{-1}$. This may imply a breaking of the perturbative expansion or a phase transition.

 A similar behaviour has been observed in superconductors  and recently discussed in \cite{Hartnoll:2008kx}\footnote{We thank P.Pani for pointing us this analogy.}. Superconductors obey the London equation
   $j_i (\omega,k) \sim A_i(\omega,k)$ that relates  the Fourier transform of the (superconducting) current density $j_i$ to the vector potential $A_i$,
with $\omega$ the frequency and $k$ the momentum. The limits $\omega\to 0$ and $ k\to 0$ do not commute: setting $\omega=0$ and then sending $k \to 0$ leads to the Meissner effect explaining the expulsion of magnetic lines out of the conductor while the opposite limit yields a current growing with  $\omega^{-1}$
explaining the superconductivity.
It would be interesting to see whether a similar physical interpretation of the two limits exists for the BH tidal response.

Finally we computed the amplification factors for BH's and ECO's in four and five dimensions that are given again in closed forms in terms of $k_0$, $p_0$, $p_2$.

\acknowledgments
We thank M. Bianchi, G. Bonelli, V. Cardoso, T. Damour, D. Fioravanti, P. Pani and A. Tanzini for useful discussions and A. Grillo for collaboration in the early stages of this work. R.Poghossian thanks the INFN Roma Tor Vergata and the framework of the Armenian SCS
grants 20RF-142 and 21AG-1C062 for financial support.
The work of F.Fucito and J.F.Morales is partially supported by the MIUR PRIN Grant 2020KR4KN2 "String Theory as a bridge between Gauge Theories and Quantum Gravity".

\begin{appendix}

\section{Instanton partition function}
\label{sec-inst}

 The instanton partition function of a four-dimensional SQGT with ${\cal N}=2$ supersymmetry, gauge group $U(2)^n$ living on an  $\Omega$-background is specified by the gauge couplings $q_i=e^{2\pi {\rm i} \tau_i}$,    $i=1,\ldots n$ being a label of the gauge group,
  the vacuum expectation values $a_{i u}$ of the scalars in the vector multiplets, $u=1,2$ a gauge index and the parameters $\epsilon_1,\epsilon_2$ specifying the $\Omega$-background. Finally we denote by  $a_{0 u}$ and  $a_{n+1, u}$ the vevs parametrizing the masses of the hypermultiplets transforming in the fundamental and anti-fundamental representations at the left and right ends of the quiver.   The quiver is displayed in figure \ref{quiv_block}.

The instanton partition function can be written as the sum \cite{Nekrasov:2002qd,Flume:2002az,Bruzzo:2002xf,Fucito:2012xc,Nekrasov:2012xe}
\be
\label{zinst}
Z^{U(2)^n}_\text{inst}= \sum_{ Y  }  \prod_{i=1}^n q_{i}^{|\vec Y_{i}|}
{  \prod_{{s}=0}^{n}    \prod_{u,v=1}^2 {\cal Z}_{Y_{s u},Y_{s+1,v} }(a_{su}-a_{s+1,v})
 \over \prod_{{i}=1}^n   \prod_{u \neq v}^2  {\cal Z}_{Y_{iu},Y_{iv} }(a_{iu}-a_{iv})}
 \ee
over an array $\vec Y_{s}=\{ Y_{{s}1},Y_{{s}2} \}$  of Young tableaux specifying the positions of the  instantons associated to the ${s}^{\rm th}$-gauge group with the tableaux at the two ends of the quiver empty, i.e. $Y_0=Y_{n+1}=\{ \emptyset, \emptyset\}$.

Each tableau $Y=\{ {k}_1\ge {k}_2\ge \cdots\} $ in (\ref{zinst})  is labelled by a sequence of decreasing integers ${k}_j$ specifying the height of the $j$-th column.  Similarly, the transpose diagram $Y^T=\{{k}'_1 \ge {k}'_2 \cdots \}$ is labelled by  ${k}'_j$ specifying the length of the $i$-th row in the original diagram. The instanton number  $|\vec Y_{s}|=\sum_u | Y_{{s}u}|$ with $|Y|$ denoting the total number of boxes in the tableau  $|Y| = \sum_{j} {k}_{j}$.
 Finally the contribution of each bifundamental is given by
\be
{\cal Z}_{Y_u,W_v}(x)
= \prod_{i,j\in Y_u}  \left[ x{+}\epsilon_1 (i{-}{k}_{jv}'){-} \epsilon_2 ( j{-}{k}_{iu}{-}1) \right]  \prod_{i,j\in W_v} \left[ x  {-}\epsilon_1 (i{-}{k}_{ju}'{-}1){+} \epsilon_2 ( j{-}{k}_{iv})   \right] \label{zyy}
\ee
  The gauge prepotential is identified with the free energy of the statistical system
 \be
 {\cal F}_{\rm inst} (a_{su},q_i.\epsilon_1,\epsilon_2)  =   \epsilon_1 \epsilon_2 \log Z_\text{inst}  (a_{su},q_i.\epsilon_1,\epsilon_2)
\ee

\section{Hypergeometric identities}

For convenience of the reader here we collect the identities used in establish braiding and fusion relations
\bea
&& {}_2F_1\left(a,b;c; z^{-1}\right)=
e^{-i \pi  a}z^a \frac{\Gamma (c) \Gamma (b-a)}{\Gamma (b) \Gamma (c-a)}
\, _2F_1(a,a-c+1;a-b+1;z) \nn \\
 &&  \qquad \qquad  +e^{-i \pi  b} z^b
\frac{(\Gamma (c) \Gamma (a-b))}{\Gamma (a) \Gamma (c-b)} \, _2F_1(b,b-c+1;-a+b+1;z)\nn\\
&& {}_2F_1\left(a,b;c;z^{-1}\right)=z^a\frac{
   \Gamma (c) \Gamma (c{-}a{-}b) \,
 }{\Gamma (c-a) \Gamma
   (c{-}b)}  {}_2F_1(a,a{-}c{+}1;a{+}b{-}c{+}1;1{-}z) \nn \\
   &&  \qquad \qquad  +e^{i \pi  (c{-}a{-}b)}
   z^a(1-z)^{c{-}a{-}b} \frac{ \Gamma (c)  \Gamma (a{+}b{-}c) \,
  }{\Gamma (a) \Gamma (b)} {}_2F_1(1{-}b,c{-}b;c{-}a{-}b+1;1{-}z) \nn\\
&& {}_2F_1\left(a,b;c;z\right)=\frac{
	\Gamma (c) \Gamma (c{-}a{-}b) \,
}{\Gamma (c-a) \Gamma(c{-}b)}  {}_2F_1(a,b;a{+}b{-}c{+}1;1{-}z) \label{hypid} \\
&&  \qquad \qquad  +(1-z)^{c{-}a{-}b} \frac{ \Gamma (c)  \Gamma (a{+}b{-}c) \,
}{\Gamma (a) \Gamma (b)} {}_2F_1(c{-}a,c{-}b;c{-}a{-}b+1;1{-}z) \nn
\eea

\end{appendix}
\providecommand{\href}[2]{#2}\begingroup\raggedright\endgroup
\end{document}